\documentclass[10pt,journal,compsoc]{IEEEtran}

\usepackage{amsfonts}
\usepackage{balance}
\usepackage{booktabs}
\usepackage{amsmath}
\usepackage{tabularx}
\usepackage{subfigure}
\usepackage{graphicx}
\usepackage{multirow}
\usepackage{rotating}
\usepackage{algorithm,algorithmic}
\usepackage{footnote}
\usepackage{threeparttable}
\usepackage{url}
\usepackage{bm}

\usepackage{color}% This package is for using color package.
\newcommand{\black}[1]{\textcolor{black}{#1}}

\ifCLASSOPTIONcompsoc
  \usepackage[nocompress]{cite}
\else
  \usepackage{cite}
\fi

\ifCLASSINFOpdf
\else
\fi

\begin{document}

\title{\black{BCFNet: A Balanced Collaborative Filtering Network with Attention Mechanism}}

\author{
	Zi-Yuan Hu,
	Jin Huang,
    Zhi-Hong Deng,
    Chang-Dong~Wang,~\IEEEmembership{Member,~IEEE,}\\
	Ling Huang,~\IEEEmembership{Student Member,~IEEE,}
    Jian-Huang~Lai,~\IEEEmembership{Senior Member,~IEEE,} and
	Philip~S.~Yu,~\IEEEmembership{Fellow,~IEEE,}
\IEEEcompsocitemizethanks{ 
\IEEEcompsocthanksitem Z.-Y. Hu, J. Huang, Z.-H. Deng, C.-D. Wang, L. Huang and J.-H. Lai are with the School of Data and Computer Science,
Sun Yat-sen University, Guangzhou, China, and also with Key Laboratory of Machine Intelligence and Advanced Computing, Ministry of Education, China.
\protect\\
E-mail: huzy33@mail2.sysu.edu.cn, huangj323@mail2.sysu.edu.cn, dengzhh7@mail2.sysu.edu.cn, changdongwang@hotmail.com, huanglinghl@hotmail.com, stsljh@mail.sysu.edu.cn.
\protect\\
\IEEEcompsocthanksitem
P. S. Yu is with University of Illinois at Chicago, Chicago, IL, U.S.A. and with Institute for Data Science, Tsinghua University, Beijing, China.
\protect\\
E-mail: psyu@cs.uic.edu.
}
}

% The paper headers
%\markboth{IEEE Transactions on Knowledge and Data Engineering}%
%{Shell \MakeLowercase{\textit{et al.}}: Bare Demo of IEEEtran.cls for Computer Society Journals}

\IEEEtitleabstractindextext{
	\begin{abstract}
		
\black{Collaborative Filtering (CF) based recommendation methods have been widely studied}, which can be generally categorized into two types, i.e., representation learning-based CF methods and matching function learning-based CF methods. Representation learning \black{\black{tries} to learn a common low dimensional space for the representations of users \black{and} items. In this case, a user and item match better if they \black{have higher similarity} in that common space.} Matching function learning \black{tries} to directly learn the complex matching function that maps user-item pairs to matching scores. Although both methods are well developed, they suffer from two fundamental flaws, i.e., \black{the representation learning \black{resorts to applying} a dot product which has limited expressiveness on the latent features of users and items, \black{while} the matching function learning has weakness in capturing low-rank relations.} \black{To overcome such flaws, we propose a \black{novel recommendation model named Balanced Collaborative Filtering Network (BCFNet)}, which has the strengths of the two types of methods. In addition, an attention mechanism is designed to better capture the hidden information within implicit feedback and strengthen the learning ability of the \black{neural network}. Furthermore, a balance module is designed to alleviate the over-fitting issue in DNNs.} Extensive experiments on \black{eight real-world} datasets demonstrate the effectiveness of the proposed \black{model}.

	\end{abstract}
	
	\begin{IEEEkeywords}
		\black{Recommender system, representation learning, matching function learning, attention mechanism, balance module.}
\end{IEEEkeywords}}

\maketitle

\IEEEdisplaynontitleabstractindextext

\IEEEpeerreviewmaketitle

\IEEEraisesectionheading{\section{Introduction}\label{sec:introduction}}

\IEEEPARstart{O}{ver} the past decades, recommender systems have been extensively studied and widely deployed in many different scenarios to alleviate the information overload problem. Due to the distinguishing capability of utilizing collective wisdom and experiences, Collaborative Filtering (CF) \black{algorithms have} been widely used to build recommender systems~\cite{Wang_Serendipitous:18,Zhao_LSCD:18,Hu_ItemRec:17,Cai2014,Wang2015rctr}.

\black{Matrix factorization is an important model in CF~\cite{Hu_Item:18}, \black{which} assumes that some relationship can be established between users and items through latent factors. By learning a common low dimensional space for the representations of users and items where they can be compared directly, the relevance of \black{users and items} can be further calculated by \black{their similarity}. In this way, \black{matrix} factorization can predict a personalized ranking for an individual user over a set of items.} \black{Unfortunately, in matrix factorization, the mapping relationship between the original representation space and the latent space is assumed to be linear, which can not be always guaranteed.} Since Deep Neural Networks (DNNs) are extremely good at representation learning of complex relationship, deep learning methods have been widely explored and have shown promising results in various areas such as computer vision, speech recognition and natural language processing~\cite{he2016deep,graves2013speech,serban2016building,Bai2019nlp}. In the past few years, there are also many works adopting DNNs \black{for recommendation and generate more accurate prediction.} \black{To better learn the complex mapping between these two spaces, Deep Structured Semantic Models (DSSM) were proposed~\cite{Huang2013dssm}, which \black{use} a deep neural network to rank for web search.} Motivated by DSSM, Xue et al.~\cite{xue2017deep} proposed a Deep Matrix Factorization (DMF), which uses a two pathways neural network architecture to replace the linear embedding operation used in vanilla matrix factorization. However, \black{\black{they still resort} to using inner product as matching function, which simply combines  the multiplication of latent features linearly and seriously limits the expressiveness of the model when \black{predicting matching scores}.}

\black{In order to learn better representation for users and items, it's a good choice to replace the dot product with a deep neural network, which can lead to better recommendation performance~\cite{xu2018deep}. DNNs are very suitable to learn the complex matching function, since they are capable of approximating any continuous function~\cite{hornik1989multilayer}.} For example, He et al.~\cite{he2017neural} proposed NeuMF under the Neural Collaborative Filtering (NCF) framework which \black{replaces the dot product operations in matrix factorization with a multi-layer neural network to capture the nonlinear relationship between users and items. \black{By taking the concatenation of user embedding and item embedding as the input of a Multi-Layer Perceptron (MLP) model, NeuMF is able to learn the interaction between users and items, from which the prediction can be made. In particular, it is capable of learning the complex mapping relationship between user-item representation and matching score. Therefore, compared} with traditional MF methods, using MLP to replace dot product on recommendation can learn a better matching function.} 

However, as revealed in~\cite{beutel2018latent}, MLP is very inefficient in catching low-rank relations. In fact, using dot product to estimate matching score in traditional matrix factorization methods is to artificially limit the model to learn similarity --- a low-rank relation that is thought to be positively related to matching score according to human experience. \black{Moreover, since lots of the training samples in recommender systems are subjected to the sparsity \black{issue}, there are only a relatively small number of ratings which can be fed into MLP. The DNNs-based model with massive parameters may easily suffer from the over-fitting issue.}

% We need a model that can more easily express and deal with multiplicative relations. That is why we incorporate MLP with a generalized matrix factorization model (GMF) in our framework. To alleviate the over-fitting issues and offset the weakness of MLP in capturing low-rank relations, we concatenate it with a balance module GMF. GMF can augment DNNs with multiplicative relation modeling. Besides, this balance module can maintain good performance even in the case of only a small number of samples. GMF applies a linear kernel to model the user-item interactions, and MLP uses a non-linear kernel to learn the interaction function. By concatenate the MLP and GMF in one model, the model is more rational because it unifies the strengths of MF and MLP. MF uses dot product to capture the multiplicative relation, while MLP is more flexible in using DNN to learn the matching function.}

According to the above discussion, we can see that there are two types of methods for implementing collaborative filtering. One is \black{mainly} based on representation learning and the other one is \black{mainly} based on matching function learning. Since these two types of methods have different advantages \black{in learning} the representation from different perspectives, a stronger and more robust joint representation for the user-item pair can be obtained by concatenating their learned representations.

%\black{In order to overcome the shortages of these two types of methods and further improve the performance of CF methods, we incorporate them and then add a balance module (GMF) and the attention mechanism under the proposed DeepCF++ framework. Besides, we mainly focus on implicit feedback, which indirectly reflect users' opinion through their behaviors. The quantity of implicit data far outweighs the quantity of explicit data in real world. }Because of that, designing recommendation algorithms that can work with implicit feedback data is extremely important and has been one of the hot research topics in recommender system.~\cite{Ding2018rs,Rendle2014,he2016vbpr} As a result, we focus on implicit feedback in this paper. 

\black{\black{In our previous work~\cite{deng2019deepcf}, we first used these two types of CF methods to obtain different representations for the input user-item pair, which are integrated together to form a Deep Collaborative Filtering (DeepCF) framework. In this paper, as an extension of DeepCF, before feeding the vectors into DNNs,} we first input them into a feed-forward attention layer which can improve the representation ability of the deep neural networks. By allowing different parts to contribute differently when compressing them to a single representation, attention-based architectures can learn to focus their ``attention'' to specific parts. Higher weights indicate that the corresponding factors are more informative for the recommendation. \black{In addition, to alleviate the over-fitting issue and offset the weakness of MLP in capturing low-rank relations, a balance module is introduced by means of generalized matrix factorization model (GMF). Therefore, \black{a novel model named Balanced Collaborative Filtering Network (BCFNet) is proposed, which consists of three sub-models, namely attentive representation learning (BCFNet-rl), attentive matching function learning (BCFNet-ml) and balance module (BCFNet-bm).}}}

%In this paper, each user is associated with a set of items. However, \black{in the case of implicit feedback}, user does not always have the equal item preferences when he/she gives a set of positive feedback. To better characterize user's preference, the implicit feedback requires different attentions on the set of items. Besides, the user who interacts with a larger number of items might have a better ability in choosing suitable items, so we can assume the recommendation model requires more attention on them among all the users. As a result, we incorporate attention mechanism into our model. 

%Extensive experiments show that the use of attention on our model can capture the hidden information within the user-item implicit interactions, and thus achieves more accurate predictions. The output from the attention layer will be fed into the two types of CF methods' deep neural network. To calculate the matching score, we then pass this joint representation into a fully connected layer which enables the model to assign different weights on the features, finally we obtain the prediction.}

The main contributions of this work are as follows.
\begin{itemize}
	\item \black{We point out the significance of incorporating collaborative filtering methods based on representation learning and matching function learning, and then \black{propose a novel BCFNet model that combines attentive representation learning, attentive matching function learning and  balance module.} The proposed \black{model} adopts the {Deep+Shallow} pattern and \black{employs attention mechanism for} collaborative filtering with implicit feedback.}
	\item \black{A feed-forward attention mechanism is utilized to better capture the hidden information within implicit feedback and strengthen the learning ability of the neural network. A balance module is also designed to alleviate the over-fitting issue caused by the high sparsity of interaction information. These two strategies enable  the proposed BCFNet model to have great flexibility in learning the complex matching function and to effectively  learn low-rank relations between users and items.}
	\item \black{Extensive experiments are conducted \black{on} eight real-world datasets to demonstrate the effectiveness and rationality of 
	\black{the BCFNet model}. The results show that the proposed \black{BCFNet} model consistently outperforms the state-of-the-art methods.} 
\end{itemize}

\black{The rest of this paper is organized as follows. Section 2 briefly reviews the related work. Section 3 is the preliminaries. \black{Section 4 introduces the \black{BCFNet model} in detail}. Section 5 presents and analyzes the \black{experimental} results. At last, Section 6 \black{draws} the conclusion of this paper.}

%\black{Some part of this work was first presented in~\cite{deng2019deepcf}.}

\section{Related Work}
\label{sec:RelatedWork}

\subsection{Collaborative Filtering with Implicit Data}
\black{\black{Compared with implicit data, it's more difficult to collect explicit feedback (e.g., product ratings)} because most of users would not tend to rate items. In fact, since users don't need to express their preference explicitly, users' implicit feedback, like a click, view times, collect or purchase history, can be more easily collected at a larger scale with a much lower cost than explicit feedback. In this case}, it's very important to design recommendation algorithms that can work with implicit feedback data~\cite{oard1998implicit,ma2013experimental}. \black{There are many well-known methods that study collaborative filtering with implicit feedback, \black{like ALS~\cite{hu2008collaborative} and SVD++~\cite{koren2008factorization}}.}

Both of the two models factorize the binary interaction matrix and assume user dislike unobserved items, i.e., assign 0 for unobserved items in the binary interaction matrix. \black{However, there are also several works considering that user may have never seen the unobserved items~\cite{rendle2009bpr,mnih2012learning,he2016vbpr}, which} tend to assuming user \black{prefers} the selected items than the unobserved ones. For example, Bayesian personalized ranking (BPR) is \black{an} effective learning algorithm for implicit CF and has been widely adopted in many related domains, which focus on \black{pair-wise} loss rather than point-wise \black{loss.}

\subsection{Collaborative Filtering based on Representation Learning}
Since Simon Funk proposed Funk-SVD~\cite{funk2006svd} in the famous Netflix Prize competition, matrix factorization for collaborative filtering has been widely studied and constantly developed over the past ten years~\cite{salakhutdinov2008bayesian,koren2009matrix,koren2009collaborative,hu2014your}. \black{\black{The main idea of these works is} mapping user and item into a common representation space where they can be compared directly.} 
Recently, deep learning methods have shown promising results in various areas such as computer vision and natural language processing. \black{Inspired by these significant success, \black{some attempts} have been made in introducing deep neural networks (DNNs) to recommender systems.  In~\cite{Wang2015cdl}, a model named Collaborative Deep Learning (CDL) was proposed, which \black{performs} deep representation learning for the content information and collaborative filtering for the rating matrix. \black{Besides, AutoRec~\cite{sedhain2015autorec}, which is the first model attempting to learn user and item representation respectively by using auto-encoder to reconstruct the input ratings, has been applied to the recommendation.} 
%\black{To overcome the limitation of auto-encoders, Collaborative Denoising Auto-Encoders (CDAE)~\cite{wu2016collaborative} further improved it by inputting both ratings and IDs.***What limitations? Why using ratings and IDs can help solve the problem***} In addition, Li et al. [47] proposed a deep architecture for collaborative filtering by combining matrix factorization with marginalized Denoising Stacked Auto-Encoders (DSAE), to overcome the sparse nature of the ratings and the side information. 
\black{In addition}, a deep learning architecture called DMF~\cite{xue2017deep} uses the rating matrix directly as the input and maps user and items into a common low-dimensional space via a deep neural network.} Overall, representation learning-based methods learn representation in different ways and can flexibly incorporate with auxiliary data. \black{However, despite their effectiveness and many subsequent developments, they still resort to \black{using} the dot product or cosine similarity as interaction function when predicting matching score.}

\subsection{Collaborative Filtering based on Matching Function Learning} 
\black{Matrix factorization (MF) has shown its \black{effectiveness in many recommender systems.} However, most of the MF methods still use dot product which limits the expressiveness of the model when doing prediction. Several recent works on neural recommender models have \black{shown that learning} the interaction function from data can obtain better recommendation prediction.} {NeuMF}~\cite{he2017neural} is a recently proposed framework that replaces the dot product used in vanilla MF with a neural network to learn the matching function. To offset the weakness of MLP in capturing low-rank relations, {NeuMF} unifies MF and MLP in one model. {NNCF}~\cite{bai2017neural} is a variant of {NeuMF} that takes user neighbors and item neighbors as inputs.  Other than {NeuMF}, there are also many other works attempting to learn the matching function directly by making full use of auxiliary data. For example, {Wide\&Deep}~\cite{cheng2016wide} adapts LR and MLP to learn the matching function from input continuous features and categorical features of user and item. {DeepFM}~\cite{guo2017deepfm} replaces LR with Factorization Machines (FM) to avoid manual feature engineering. \black{Neural Factorization Machines (NFM)~\cite{he2017nfm} uses} a bi-interaction pooling layer to learn feature crosses. What's more, tree-based models are also studied and proven to be effective~\cite{zhao2017gb,zhu2017deep,wang2018tem}. \black{Neural network based Aspect-level Collaborative Filtering model (NeuACF) has been applied to exploit different aspect latent
factors by using attention mechanism with NCF~\cite{Shi2019NeuACF}. \black{ConvNCF~\cite{he2018outer} uses an outer product operation to replace concatenation used in {NeuMF} and utilizes 2D convolution layers for learning joint representation of user-item pairs.} In this paper, we mainly} focus on pure collaborative filtering without using auxiliary data.

\subsection{\black{Attention Mechanism in Recommender System}}
\black{Attention mechanism \black{has shown effectiveness} in various machine learning tasks such as machine translation and computer \black{vision~\cite{Bahdanau2015nmt}.} Recently, several works have done in utilizing attention mechanism in  \black{recommender systems~\cite{He2018NAIS,Shi2019NeuACF,tay2018self,Cao2019seagr,Xi2019bpam}.} For instance, in~\cite{Chen2017acf}, a model named Attentive Collaborative Filtering (ACF) was proposed to employ attention modeling in CF. In~\cite{Cheng2018a3ncf}, the attention mechanism was introduced to capture the varying attention vectors of each specific user-item pair. To improve the performance of factorization, a model named attention factorization machine \black{for learning} the weight of feature interactions via attention networks was designed\black{~\cite{Xiao2017afm}}. \black{In~\cite{Zhou2017atrank}}, an attention-based user model called ATRank was proposed, which \black{utilizes} a novel attention mechanism to model user behavior after considering the influences brought by other behaviors. Attention mechanism \black{derives} from the idea that human recognition usually can not process the whole entire signal at once, instead, one only focuses on few selective parts at a time. Attention's success is mainly due to its advantage \black{in assigning} attentive weights for the input vectors, \black{where} higher weights indicate that the corresponding factors are more informative for the recommendation. In this paper, we adopt the attention mechanism in our model to make more accurate predictions.}

According to the above discussion, both representation learning-based and matching function learning-based collaborative filtering methods have been broadly studied and proven to be effective. Despite their strengths, both of the two types of methods have weaknesses, i.e., the limit expressiveness of dot product and the weakness in capturing low-rank relations. \black{In our previous work~\cite{deng2019deepcf}, we pointed out the significance of combining the two types of collaborative filtering methods to overcome these weaknesses.} \black{In this paper we present a \black{novel model} that ensembles these two types of methods, \black{and} add a balance module and the attention mechanism to endow the model with a great flexibility of learning the matching function while maintaining the ability to learn low-rank relations efficiently. For clarity, \black{Table}~\ref{table:notations} summarizes the main notations used in this paper.}

\begin{table}[!t]
	\caption{\black{Summary of the main notations.}}
	\label{table:notations}
	\centering{
		%\begin{center}
		\begin{tabular}{@{}c@{}|l@{}}
			\toprule
			${M}$                	&The number of users\\
			\hline
			${N}$                	&The number of items\\
			\hline
			$\mathbf{Y}=[y_{u,i}] \in \mathbb{R}^{M \times N}$ & Binary user-item interaction matrix\\
			\hline
			$y_{u,i}$			     	&The interaction of user $u$ to item $i$ \\
			\hline
			${\rho}$  & Negative sample ratio\\
			\hline
			$\mathcal{Y}^+$  & All the observed interactions in $\mathbf{Y}$\\
			\hline
			$\mathcal{Y}^-$  & All the unobserved interactions in $\mathbf{Y}$\\
			\hline
			$p_{u,i}$              &The probability that user $u$ is matched by item $i$\\
			\hline
			$\hat{y}_{u,i}$			&The predicted interaction of user $u$ to item $i$\\
			%\hline
			%$f$ 					&The mapping function of the model\\%Wang marks: We don't need to list this function as $f$ is used many times with different meanings.
			\hline
			$\Theta$ 				&The model parameters\\
			\hline
			$\mathbf{v}_u^U$   &The initial representation of user $u$\\
			\hline
			$\mathbf{v}_i^I$   &The initial representation of item $i$\\
			\hline
			$\mathbf{p}_u$   &The latent representation of user $u$\\
			\hline
			$\mathbf{q}_i$   &The latent representation of item $i$\\
			\hline
			\multirow{2}*{$\mathbf{v}_e$}   &The encoder vector of the feed-forward\\& attention layer\\
			\hline
			\multirow{2}*{$\mathbf{v}_d$}   &The decoder vector of the feed-forward\\& attention layer\\
			\hline
			\multirow{2}*{$\bm{\alpha}$}   &The attention ratio of the feed-forward\\& attention layer\\
			\hline
			$\mathbf{W}$   &Weight matrix\\
			\hline
			$\mathbf{b}$   &Bias vector\\
			\hline
			$\mathbf{a}_\mathrm{Y}^{rl}$   &The predictive vector of BCFNet-rl\\
			\hline
			$\mathbf{a}_\mathrm{Y}^{ml}$   &The predictive vector of BCFNet-ml\\
			\hline
			$\mathbf{a}_\mathrm{Y}^{bm}$   &The predictive vector of BCFNet-bm\\
			\bottomrule
		\end{tabular}
	}
	%\end{center}
\end{table}

\section{Preliminaries}
\subsection{Implicit Feedback Data}
Suppose there are $M$ users and $N$ items in the system, following~\cite{wu2016collaborative,he2017neural}, we construct the user-item interaction matrix \black{$\mathbf{Y}=[y_{ui}]\in \mathbb{R}^{M \times N}$} from users' implicit feedback as follows,
\begin{align}\label{eq:interaction_matrix}
y_{ui} = 
\begin{cases}
1,& \text{if interaction (user } u \text{, item } i\text{) is observed;}\\
0,& \text{otherwise.}
\end{cases}
\end{align}

\black{Although \black{compared} with explicit feedback, implicit feedback can be easier to obtain, it is more challenging to be utilized because it has two major problems. First, unlike ratings, implicit feedback is inherently noisy. While we track \black{a} user-item interaction ($y_{ui} = 1$), we can only guess users' preference indirectly\black{. For} example,  the observed interaction does not provide any specific information about how much exactly a user likes an item.} Second, \black{without an observed} interaction ($y_{ui} = 0$) does not mean user $u$ does not like item $i$. In fact, user $u$ may have never seen item $i$ since there are too many items in a system. \black{The non-observed user-item interactions may be a mixture of real negative feedback and missing values.} These two problems pose huge challenges in learning from implicit data, especially the second one. 

To perform collaborative filtering on implicit data which lacks real negative feedback is also known as the One-Class Collaborative Filtering (OCCF) problem~\cite{pan2008one}. \black{To tackle the problem of unobserved negative samples, several approaches have been proposed which can be classified into two categories: whole data based learning and sample based learning. The former assumes that all the unobserved data are weak negative instances and are equally weighted~\cite{hu2008collaborative,pan2008one}}, \black{while the latter samples some negative instances from unobserved interactions~\cite{pan2008one,wu2016collaborative,he2017neural}. In this paper, we perform the second method, i.e., uniformly sample negative instances from unobserved interactions} \black{with the negative sample ratio $\rho$, i.e., the number of negative samples per positive instance}. \black{Later we will conduct some experiments to verify the impact of $\rho$ on the proposed model.} \black{Let $\mathcal{Y}^+$ denote all the observed interactions in $\mathbf{Y}$ and $\mathcal{Y}^-$ denote the sampled unobserved interactions, i.e., the negative instances.} 

%The recommendation problem with explicit feedback is usually formulated as a rating prediction problem which estimates the missing values in rating matrix $\mathbf{R}$. The predicted scores are then used for ranking items and finally the top-ranking items are recommended to users\black{, which is called the top-K recommendation problem.} Similarly, t

\black{To tackle the recommendation problem with implicit feedback}, we can formulate it as an interaction prediction problem which estimates the missing values in interaction matrix $\mathbf{Y}$, i.e., estimates whether the unobserved interactions would happen or not \black{(the user would give a rating on the item or not)}. However, unlike explicit feedback, implicit feedback is discrete and binary. \black{When \black{dealing} with implicit feedback that each entry is a binary value of 1 or 0, we often consider the learning of a recommender model as a binary classification problem.} Solving the above binary classification problem can not help us to further rank and recommend items. One feasible solution is to employ a probabilistic treatment for interaction matrix $\mathbf{Y}$. We can assume $y_{ui}$ obeys a Bernoulli distribution:
\begin{align}
\label{eq:bernoulli_distribution}
\nonumber P(y_{ui} = k \vert p_{ui}) =&
\begin{cases}
1-p_{ui},& k=0;\\
p_{ui},& k=1
\end{cases}\\
=& p_{ui}^k(1-p_{ui})^{1-k},
\end{align}
where $p_{ui}$ is the probability of $y_{ui}$ being equal to 1. What's more, $p_{ui}$ can be further interpreted as the probability that user $u$ is matched by item $i$. In this case, a value of 1 for $p_{ui}$ indicates that item $i$ perfectly matches user $u$ and a value of 0 indicates that user $u$ and item $i$ do not match at all. Rather than modeling $y_{ui}$ which is discrete and binary, our method models $p_{ui}$ instead. In this manner, we transform the binary classification problem, i.e., the interaction prediction problem, to a matching score prediction problem.

\subsection{\black{Feed-forward Attention Layer}}
\black{In order to enhance the learning ability of \black{Deep Neural Networks (DNNs)}, we utilize a feed-forward attention mechanism~\cite{Bahdanau2015nmt,raffel2015feed} before the learning process. Attention mechanism \black{has been shown to be effective} in various machine learning tasks such as machine translation, recommendation and computer vision~\cite{Zhou2018DIN,Xiao2017afm}. It has the advantage that it \black{can} assign different attentive scores \black{to} the input vectors, \black{with higher values indicating} that the corresponding vectors are more informative. For the representation learning-based CF methods, a feed-forward attention layer can be added before the representation function to enhance its learning ability for user features and item features respectively. \black{Similarly}, for the matching function learning-based CF methods, we also can add a feed-forward attention layer before its matching function, which can also improve its learning \black{performance}.}

\black{Suppose that there is a $m$-dimensional vector $\mathbf{v}_e$ as the input of a feed-forward attention layer\black{, which is called the encoder vector.} Then the output of the attention layer is also a $m$-dimensional vector $\mathbf{v}_d$\black{, which is called the decoder vector.} In this paper, we adopt a BP neural network~\cite{goh1995back} to learn the relationship between $\mathbf{v}_e$ and $\mathbf{v}_d$. Therefore, the calculation process of $\mathbf{v}_d$ can be formulated as:
	\begin{align}\label{eq:ffal}
	\begin{split}
	\black{\bm{\alpha}} & = \delta(\mathbf{W}^T\mathbf{v}_e+\mathbf{b})\\
	\mathbf{v}_d&=\black{\bm{\alpha}} \odot \mathbf{v}_e,
	\end{split}
	\end{align}
where $\delta$ denotes the activation function $Softmax$, $\mathbf{W}\in\mathbb{R}^{m\times m}$ and $\mathbf{b}\in\mathbb{R}^{m\times 1}$ \black{denote} the weight matrix and bias vector of the BP neural network, $\black{\bm{\alpha}}\in\mathbb{R}^{m\times 1}$ denotes the attention ratio of the feed-forward attention layer\black{, and $\odot$ denotes the element-wise product of $\black{\bm{\alpha}}$ and $\mathbf{v}_e$}. We utilize a \black{$Softmax$} function as the activation function to obtain the attention ratio $\black{\bm{\alpha}}$, \black{i.e., the} probability vector by the BP neural network. The decoder vector $\mathbf{v}_d$ is ultimately calculated by $\black{\bm{\alpha}}$ and $\mathbf{v}_e$, which can reflect the importance of each element in $\mathbf{v}_e$. In this formulation, attention mechanism can be \black{regarded} as computing an adaptive \black{weight} of the encoder vector $\mathbf{v}_e$. And then the decoder vector $\mathbf{v}_d$ can be used for representation learning and matching function learning. We \black{will} also conduct some experiments to \black{demonstrate} that the feed-forward attention mechanism can improve the learning ability of our model.} 

\subsection{Learning the Model}

A model-based method generally assumes that data can be generated by an underlying model as $\hat{y}_{ui} = f(u, i \vert \Theta)$, where $\hat{y}_{ui}$ denotes the prediction of $y_{ui}$, i.e., the predicted probability that user $u$ is matched by item $i$, $\Theta$ denotes model parameters, and $f$ denotes the function that maps model parameters to the predicted score. In this manner, we need to figure out two key questions, i.e., how to define function $f$ and how to estimate parameters $\Theta$. We will answer the first question in the next section.

For the second question, \black{to estimate parameters $\Theta$, most of the existing works generally optimize an objective function.} Two types of objective functions are commonly used in recommender system, namely, {point-wise loss}~\cite{hu2008collaborative,he2016fast} and {pair-wise loss}~\cite{rendle2009bpr,mnih2012learning,he2016vbpr}. \black{Point-wise loss learning methods usually try to minimize the loss between $\hat{y}_{ui}$ and its target value ${y}_{ui}$, while the pair-wise learning maximizes the margin between observed entry $\hat{y}_{ui}$ and unobserved entry ${y}_{ui}$.} In this paper, we explore the point-wise loss only and leave the pair-wise loss in our future work. Point-wise loss has been widely studied in collaborative filtering with explicit feedback under the regression framework~\cite{funk2006svd,salakhutdinov2008bayesian}. \black{Existing point-wise methods \black{usually}  perform a regression with squared loss (SE) to learn the recommender model. However, the squared loss can be derived by assuming that the error between \black{the given rating and the predicted rating} is generated from a Gaussian distribution, which does not hold in the implicit feedback scenario since $y_{ui}$ is discrete and binary. Thus we point out that it \black{may be unsuitable} for implicit data.} \black{As aforementioned}, \black{to adapt the binary and discrete characters of the implicit feedback data\black{,}} we assume $y_{ui}$ obeys a Bernoulli distribution, i.e., $y_{ui} \sim  Bern(p_{ui})$. By replacing $p_{ui}$ with $\hat{y}_{ui}$ in \black{Eq.~(\ref{eq:bernoulli_distribution})}, we can define the likelihood function as
\begin{align}\label{eq:likelihood_function}
\begin{split}
L(\Theta)
& = \prod_{(u,i) \in \mathcal{Y}^+ \cup \mathcal{Y}^-} P(y_{ui} \vert \Theta )\\
& = \prod_{(u,i) \in \mathcal{Y}^+ \cup \mathcal{Y}^-} \hat{y}_{ui}^{y_{ui}}(1-\hat{y}_{ui})^{1-y_{ui}},
\end{split}
\end{align}
where $\mathcal{Y}^+$ denotes all the observed interactions in $\mathbf{Y}$ and $\mathcal{Y}^-$ denotes the sampled unobserved interactions, i.e., the negative instances. Furthermore, taking the negative logarithm of the likelihood (NLL), we obtain
\begin{align}\label{eq:NLL}
\ell_{BCE}=-\sum_{(u,i) \in \mathcal{Y}^+ \cup \mathcal{Y}^-} y_{ui}\log\hat{y}_{ui} + (1-y_{ui})\log(1-\hat{y}_{ui}).
\end{align}

Based on all the above assumptions and formulations, we finally obtain an objective function which is suitable for learning from implicit feedback data, i.e., the binary cross-entropy loss function~\cite{james2013cross}. \black{By adapting a gradient descent we can optimize the objective function and minimize it for the \black{BCFNet} model.}

To sum up, the recommendation problem with implicit feedback can be formulated as an interaction prediction problem. \black{To endow our algorithm with the ability to rank items \black{for the recommendation task}, we need to employ a probabilistic treatment for interaction matrix $\mathbf{Y}$. We use a Logistic function as the activation function for the output layer, so that $y_{ui}$ is constrained in the range of [0,1].} Instead of modeling $y_{ui}$, we model $p_{ui}$ which is the probability of $y_{ui}$ being equal to 1. Since $p_{ui}$ can also be interpreted as the probability that user $u$ is matched by item $i$, the interaction prediction problem can be transformed to a matching score prediction problem. In this manner, using maximum likelihood estimation to estimate model parameters $\Theta$ is equivalent to minimizing the binary cross-entropy between $y_{ui}$ and $\hat{y}_{ui}$. 

\begin{figure*}[!t]
	\centering
	\includegraphics[width=1.0\linewidth]{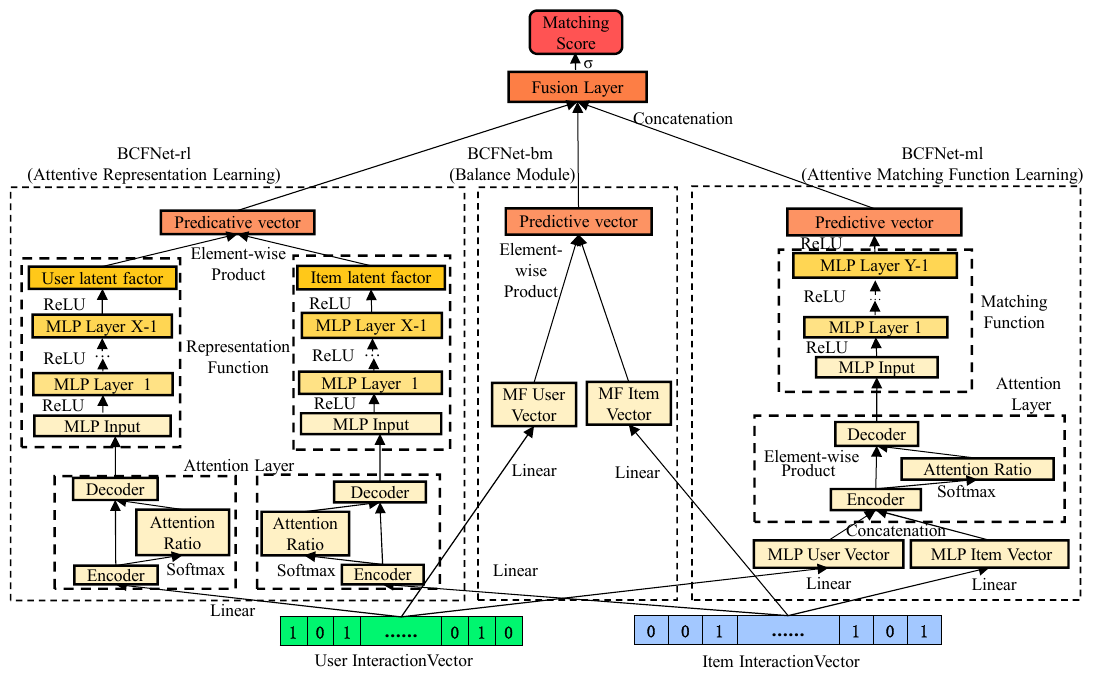}
	\caption{The architecture of the proposed \black{BCFNet} model.}
	\label{fig:architecture}
\end{figure*}

\section{The Proposed \black {Model}}
\label{sec:theproposedframework}
\black{In this section, we will \black{introduce} \black{the proposed model named Balanced Collaborative Filtering Network (BCFNet) in detail.} First, we \black {present an architecture of of the BCFNet model.} Then, we respectively introduce three sub-modules of the \black{model}, namely \black{attentive representation learning (BCFNet-rl)}, \black{attentive matching function learning (BCFNet-ml)} and balance module (BCFNet-bm)\black{. Finally, we describe} how to fuse these three sub-modules and how to learn the final BCFNet model.}

\subsection{\black{The Architecture of BCFNet}}
\black{The proposed \black{BCFNet model} consists of three sub-modules\black{, namely, attentive representation learning}, \black{attentive matching function learning} and balance module. \black{The architecture of the BCFNet model is shown in \figurename~\ref{fig:architecture}}.}

\black{
All of the three \black{modules} start from extracting data from database. IDs, historical behaviors and other auxiliary data
can all be used to construct the initial representations of user $u$ and item $i$, which are denoted by $\mathbf{v}_u^U$ and $\mathbf{v}_i^I$ respectively. The CF models then calculate $\mathbf{p}_u = f(\mathbf{v}_u^U)$ and $\mathbf{q}_i = g(\mathbf{v}_i^I)$, i.e., the latent representations for user $u$ and item $i$. Next, a non-parametric operation is performed on $\mathbf{p}_u$ and $\mathbf{q}_i$ to aggregate the latent representations. Finally, mapping function $h(\cdot)$ is used to calculate the matching score $\hat{y}_{ui}$. In \black{what follows}, we will introduce these three sub-modules and their implementations in detail. 
}

\subsection{\black{Attentive Representation Learning}}
For representation learning-based CF methods, the model focuses more on learning representation function and the matching function is usually assumed to be simple and non-parametric, e.g., dot product or cosine similarity. In this manner, the model is supposed to \black{map} users and items into a common space where they can be directly compared. For example, taking one-hot IDs as inputs, the vanilla MF~\cite{funk2006svd} adopts linear embedding function as function $f(\cdot)$ and function $g(\cdot)$ to learn the latent representations. The latent representations $\mathbf{p}_u$ and $\mathbf{q}_i$ are then aggregated by the dot product function to calculate the matching score. In this case, mapping function $h(\cdot)$ is assumed to be the identity function. For another example, taking ratings as inputs, DMF~\cite{xue2017deep} adopts MLP as function $f(\cdot)$ and function $g(\cdot)$ to learn better latent representation by making full use of the non-linearity and high capacity characteristics of neural networks. The cosine similarity between $\mathbf{p}_u$ and $\mathbf{q}_i$ is then calculated and used as matching score.

\black{\black{In this paper}, we focus on implicit feedback data only in the BCFNet model, so no auxiliary data are used. In particular, the user-item interaction matrix $\mathbf{Y}$ is taken as input. That is, the initial user representation of user $u$ is the $u$-th row vector of $\mathbf{Y}$, i.e. $\mathbf{v}_u^U=\mathbf{Y}_{u*}$, and the initial item representation of item $i$ is the $i$-th column vector of $\mathbf{Y}$, i.e. $\mathbf{v}_i^I=\mathbf{Y}_{*i}$.}

\black{In the proposed \black{BCFNet model}, we design an attentive representation learning-based CF method that combines MLP with feed-forward attention to learn latent representations for users and items.} \black{Suppose the size of the encoder layer of the feed-forward attention layer is $l$. First, for user $u$, the decoder vector $\mathbf{v}_d\in\mathbb{R}^{l\times 1}$ can be computed by applying the feed-forward attention layer to $\mathbf{Y}_{u*}\in\mathbb{R}^{\black{1\times N}}$  as follows:
	\begin{align}\label{eq:rlffal}
	\begin{split}
	\mathbf{a}_0 & = \black{\mathbf{W}_0^T\mathbf{Y}_{u*}^T}\\
	\black{\bm{\alpha}} & = \delta(\mathbf{W}^T\mathbf{a}_0+\mathbf{b})\\
	\mathbf{v}_d&=\black{\bm{\alpha}} \odot \mathbf{a}_0,
	\end{split}
	\end{align}
where $\mathbf{W}_0\in\mathbb{R}^{N\times l}$ denotes the weight matrix of the encoder layer of the feed-forward attention layer, $a_0\in\mathbb{R}^{l\times 1}$ denotes the activation of the input layer, $\delta$ denotes the activation function $Softmax$, $\mathbf{W}\in\mathbb{R}^{l\times l}$ and $\mathbf{b}\in\mathbb{R}^{l\times 1}$ \black{denote} the weight matrix and bias vector of the BP neural network respectively, and $\black{\bm{\alpha}}\in\mathbb{R}^{l\times 1}$ denotes the attention ratio of the feed-forward attention layer. And then,} the representation learning part \black{based on MLP implementation} for users can be defined as:
\black{
	\begin{align}\label{eq:rlrl}
	\begin{split}
	\mathbf{a}_0' &= \begin{bmatrix}\mathbf{a}_0\\\mathbf{v}_d\end{bmatrix}\\
	\mathbf{a}_1
	& = \black{\theta}(\mathbf{W}_1^T\mathbf{a}_0' + \mathbf{b}_1)\\
	\mathbf{a}_2
	& = \black{\theta}(\mathbf{W}_2^T\mathbf{a}_1 + \mathbf{b}_2)\\
	& \cdot\cdot\cdot\cdot\cdot\cdot\\
	\mathbf{p}_u = \mathbf{a}_\mathrm{X}
	& = \black{\theta}(\mathbf{W}_{X}^T\mathbf{a}_\mathrm{{X}-1} + \mathbf{b}_\mathrm{X}),
	\end{split}
	\end{align}
where $a_0'\in\mathbb{R}^{2l\times 1}$ \black{denotes} the input of \black{MLP},} $\mathbf{W}_x$, $\mathbf{b}_x$, and $\mathbf{a}_x$ denote the weight matrix, bias vector and activation for the $x$-th layer's perceptron respectively. $\black{\theta}(\cdot)$ is the activation function and we use $ReLU$ function in this paper. The latent representation $\mathbf{q}_i$ for item $i$ is calculated in the same manner. Different from the existing representation learning-based CF methods, the matching function part is defined as:
\black{
	\begin{align}\label{eq:rlml}
	\begin{split}
	\mathbf{a}_\mathrm Y^{rl}&=\mathbf{p}_u \odot \mathbf{q}_i\\
	\hat{y}_{ui} &= \sigma(\mathbf{W}_{out}^T\mathbf{a}_\mathrm Y^{rl}),\\
	\end{split}
	\end{align}
}where $\mathbf{a}_\mathrm Y^{rl}$, $\mathbf{W}_{out}$ and $\sigma(\cdot)$ denote \black{the predictive vector,} the weight matrix and \black{the activation function $Sigmoid$ respectively}. By substituting the non-parametric {dot product or} cosine similarity with element-wise product and a parametric neural network layer, our model still focuses on catching low-rank relations between users and items but is more expressive since the importance of latent dimensions can be different and the mapping can be non-linear. 

In summary, the \black{attentive} representation learning \black{module} used in this paper is implemented by \black{Equations~(\ref{eq:rlffal}),~(\ref{eq:rlrl}) and~(\ref{eq:rlml}}), which is called BCFNet-rl. \black{The \black{main} architecture of BCFNet-rl is shown in \black{the left dashed box of \figurename~\ref{fig:architecture}, except that the final prediction step for $\hat{y}_{ui}$ is ignored.}}

\subsection{\black{Attentive Matching Function Learning}}
\black{Matching} function learning-based CF methods focus more on matching function learning. The representation learning part is still necessary since \black{the initial representations of users and items, namely $\mathbf{v}_u^U$ and $\mathbf{v}_i^I$} are usually extremely sparse and have high dimension, making it difficult for the model to directly learn the matching function. Therefore, matching function learning-based CF methods usually use a linear embedding layer to learn latent representations for users and items. With the dense low-dimensional latent representations, the model is able to learn the matching function more efficiently.

\black{In the proposed \black{BCFNet model}, we design an attentive matching function learning-based CF method that combines MLP with feed-forward attention} to learn the matching function. Instead of IDs, we take the interaction matrix $\mathbf{Y}$ as input. \black{Suppose the size of the embedding layer is \black{$h$}. First, for user $u$ and item $i$, \black{the decoder vector $\mathbf{v}_d\in\mathbb{R}^{2\black{h}\times 1}$} can be computed by applying the feed-forward attention layer to $\mathbf{Y}_{u*}\in\mathbb{R}^{\black{1\times N}}$ and $\mathbf{Y}_{*i}\in\mathbb{R}^{M\times 1}$ as follows:
	\begin{align}\label{eq:mlffal}
	\begin{split}
	\mathbf{p}_u
	& = \black{\mathbf{P}^T\mathbf{Y}_{u*}^T}\\
	\mathbf{q}_i
	& = \mathbf{Q}^T\mathbf{Y}_{*i}\\
	\mathbf{a}_0
	& = \begin{bmatrix}\mathbf{p}_u\\\mathbf{q}_i\end{bmatrix}\\
	\black{\bm{\alpha}} & = \delta(\mathbf{W}^T\mathbf{a}_0+\mathbf{b})\\
	\mathbf{v}_d&=\black{\bm{\alpha}} \odot \mathbf{a}_0,
	\end{split}
	\end{align}
}where \black{$\mathbf{P}\in\mathbb{R}^{N\times \black{h}}$ and $\mathbf{Q}\in\mathbb{R}^{M\times \black{h}}$ are the parameter matrices of the linear embedding layers, $\mathbf{p}_u\in\mathbb{R}^{\black{h}\times 1}$ denotes the MLP user vector, and $\mathbf{q}_i\in\mathbb{R}^{\black{h}\times 1}$ denotes the MLP item vector.} \black{The matching function learning part based on MLP implementation can be defined as:
	\begin{align}\label{eq:ml}
	\begin{split}
	\mathbf{a}_0' &= \begin{bmatrix}\mathbf{a}_0\\\mathbf{v}_d\end{bmatrix}\\
	\mathbf{a}_1
	& = \black{\theta}(\mathbf{W}_1^T\mathbf{a}_0' + \mathbf{b}_1)\\
	\mathbf{a}_2
	& = \black{\theta}(\mathbf{W}_2^T\mathbf{a}_1 + \mathbf{b}_2)\\
	& \cdot\cdot\cdot\cdot\cdot\cdot\\
	\mathbf{a}_\mathrm Y^{ml}=\mathbf{a}_\mathrm{Y}
	&= \black{\theta}(\mathbf{W}_\mathrm{Y}^T\mathbf{a}_\mathrm{{Y}-1} + \mathbf{b}_\mathrm{Y})\\
	\hat{y}_{ui} 
	& = \sigma(\mathbf{W}_{out}^T\mathbf{a}_\mathrm Y^{ml}),
	\end{split}
	\end{align}
where \black{$a_0'\in\mathbb{R}^{4\black{h}\times 1}$ \black{denotes} the input of \black{MLP},} $\mathbf{W}_x$, $\mathbf{b}_x$, and $\mathbf{a}_x$ denote the weight matrix, bias vector and activation for the $x$-th layer's perceptron respectively, and $\mathbf{a}_\mathrm Y^{ml}$ denotes the predictive vector.} In this manner, the \black{attentive} representation learning functions $f(\cdot)$ and $g(\cdot)$ are implemented by the linear embedding layers. The latent representations $\mathbf{p}_u$ and $\mathbf{q}_i$ are then aggregated by a simple concatenation operation. After the process of the feed-forward attention layer, MLP is used as the mapping function $h(\cdot)$ to calculate the matching score $\hat{y}_{ui}$. Notice that although concatenation is the simplest aggregation operation, it maintains maximally the information passed from the previous layer and allows to make full use of the flexibility of the MLP model.

In summary, the \black{attentive} matching function learning \black{module} used in this paper is implemented by \black{Equations~(\ref{eq:mlffal}) and~(\ref{eq:ml})}, which is called BCFNet-ml. \black{The \black{main} architecture of BCFNet-ml is shown in \black{the right dashed box of \figurename~\ref{fig:architecture}, except that the final prediction step for $\hat{y}_{ui}$ is ignored.}}

\subsection{Balance Module}
\black{\black{After additionally introducing the attention mechanism, the unified framework of representation learning and matching function learning (i.e. the DeepCF framework proposed in the previous version~\cite{deng2019deepcf})} can be greatly improved. However, due to its DNNs structure, it may also lead to partial information loss and over-fitting issue. For a real-life \black{recommender} system, there exist \black{a large number of users and items} which are subjected to the sparsity problems. In this case, only relatively few interactions can be input into the MLP implements, which is prone to over-fitting issue and leads to mediocre results. In addition, during the deep learning process, some features of users and items may be simply ignored and some important implicit feedback may be given low weight \black{in MLP.}
}

\black{
	Inspired by some shallow recommendation model without neural networks and attention mechanism, we add the \black{generalized matrix factorization (GMF)} model~\cite{xue2017deep} to \black{the BCFNet model} as a balance module. As a shallow matrix factorization model, GMF adopts linear embedding function as representation function and \black{uses} dot product as matching function, which can offset the weakness of MLP in capturing low-rank relations and alleviate the over-fitting \black{issue} in DNNs. Therefore, assuming the size of the embedding layer is $r$, the balance module can be formulated as: 
	\begin{align}\label{eq:bm}
	\begin{split}
	\mathbf{p}_u
	& = \black{\mathbf{P}^T\mathbf{Y}_{u*}^T}\\
	\mathbf{q}_i
	& = \mathbf{Q}^T\mathbf{Y}_{*i}\\
	\mathbf{a}_\mathrm Y^{bm}&=\mathbf{p}_u \odot \mathbf{q}_i\\
	\hat{y}_{ui} &= \sigma(\mathbf{W}_{out}^T\mathbf{a}_\mathrm Y^{bm}),\\
	\end{split}
	\end{align}
where $\mathbf{P}\in\mathbb{R}^{N\times r}$ and $\mathbf{Q}\in\mathbb{R}^{M\times r}$ are the parameter matrices of the linear embedding layers, $\mathbf{p}_u\in\mathbb{R}^{r\times 1}$ denotes the MF user vector, $\mathbf{q}_i\in\mathbb{R}^{r\times 1}$ denotes the MF item vector, and $\mathbf{a}_\mathrm Y^{bm}$ denotes the predictive vector. In the following experiments, it will be verified that the balance module is helpful to alleviate the over-fitting issue caused by the high sparsity of \black{interaction} information.}

\black{
	In summary, the \black{balance} module used in this paper is implemented by \black{Eq.~(\ref{eq:bm})}, which is called BCFNet-bm. The \black{main} architecture of BCFNet-bm is shown \black{in the middle dashed box of \figurename~\ref{fig:architecture}, except that the final prediction step for $\hat{y}_{ui}$ is ignored.}}

\subsection{Fusion and Learning}

\subsubsection{Fusion} 
In the previous three subsections, \black{we have introduced the three modules of the proposed \black{BCFNet model}, each of which can be regarded as a separate model for recommender system.} To incorporate these three \black{modules}, we need to design a strategy to fuse them so that they can enhance each other \black{and improve the accuracy of the recommendation system}. One of the most common fusing strategies is to concatenate the learned representations to obtain a joint representation and then feed it into a fully connected layer. \black{As described in the previous three subsections, for BCFNet-rl, BCFNet-ml and BCFNet-bm, they generate the predictive vectors respectively, which are denoted as $\mathbf{a}_\mathrm Y^{rl}$, $\mathbf{a}_\mathrm Y^{ml}$ and $\mathbf{a}_\mathrm Y^{bm}$. And} the predictive vectors can be viewed as the representation for the corresponding user-item pair. Since the three types of CF methods have different advantages and learn the predictive vectors from different perspectives, the concatenation of the three predictive vectors will result in a stronger and more robust joint representation for the user-item pair. What's more, the consequent fully connected layer enables the model to assign different weights on the features contained in the joint representation. \black{\black{Therefore, the} output of the fusion model can be defined as:
	\begin{align}\label{eq:output}
	\hat{y}_{ui} = \sigma(\mathbf{W}_{out}^T\begin{bmatrix}\mathbf{a}_\mathrm Y^{rl}\\\mathbf{a}_\mathrm Y^{bm}\\\mathbf{a}_\mathrm Y^{ml}\end{bmatrix}).
	\end{align}
}

\black{
Using \black{Eq.~(\ref{eq:output})} to incorporate BCFNet-rl, BCFNet-ml and BCFNet-bm, we finally obtain the proposed BCFNet model.}

\begin{algorithm}[!t]
	\renewcommand{\algorithmicrequire}{\textbf{Input:}}
	\renewcommand{\algorithmicensure}{\textbf{Output:}}
	\caption{The learning algorithm for the proposed BCFNet model without pre-training.}
	\label{alg:1}
	\textbf{Input:} $\mathbf{Y}$: user-item interaction matrix; $\mathcal{Y}^+$: all the observed interactions in $\mathbf{Y}$; $\mathcal{Y}^-$: the sampled unobserved interactions; n: number of epochs. \\
	\textbf{Output:} $\mathrm{\Theta}$: the model parameters.
	\begin{algorithmic}[1]
		\STATE Randomly initialize the model parameters $\mathrm{\Theta}$  with a Gaussian distribution.
		%\REPEAT
		\FORALL{$epochs = 1$ to n}
		\FORALL{$(u,i) \in \mathcal{Y}^+ \cup \mathcal{Y}^-$}
		\STATE Compute the predictive vector of BCFNet-rl $\mathbf{a}_\mathrm Y^{rl}$ via Eq.~(\ref{eq:rlffal}), Eq.~(\ref{eq:rlrl}) and Eq.~(\ref{eq:rlml})
		\STATE Compute the predictive vector of BCFNet-ml $\mathbf{a}_\mathrm Y^{ml}$ via Eq.~(\ref{eq:mlffal}) and  Eq.~(\ref{eq:ml})
		\STATE Compute the predictive vector of BCFNet-bm $\mathbf{a}_\mathrm Y^{bm}$ via Eq.~(\ref{eq:bm})
		\STATE Obtain the predicted interaction $\hat{y}_{u,i}$ via Eq.~(\ref{eq:output})
		\ENDFOR
		\STATE Obtain the loss $\ell_{BCE}$ via Eq.~(\ref{eq:NLL})
		\STATE Use mini-batch Adam to optimize $\ell_{BCE}$
		\ENDFOR
		%\STATE 
		%\UNTIL{Eq.~(\ref{eq:NLL}) converges}
		\STATE \textbf{Return} $\mathrm{\Theta}$
	\end{algorithmic}  
\end{algorithm}

\subsubsection{Learning} 
As discussed in the previous section, \black{the objective function for the \black{BCFNet model}} is the binary cross-entropy function. To optimize the model, we use mini-batch Adam~\cite{kingma2014adam}. \black{The batch size is fixed to 256 and the learning rate is 0.00001}. The model parameters are randomly initialized with a Gaussian distribution (with a mean of 0 and standard deviation of 0.01) and the negative instances $\mathcal{Y}^-$ are uniformly sampled from unobserved interactions in each iteration. \black{The learning algorithm for the proposed BCFNet model is summarized in Algorithm~\ref{alg:1}.}

\subsubsection{Pre-training} According to~\cite{erhan2010does}, the initialization is of significance to the convergence and performance of deep learning model. Using pre-trained models to initialize the ensemble model can significantly increase the convergence speed and improve the final performance. \black{Since BCFNet is composed of three components, i.e., BCFNet-rl, BCFNet-ml and BCFNet-bm, we can pre-train these three components and use them to initialize BCFNet.} Notice that BCFNet-rl, BCFNet-ml and BCFNet-bm are trained from scratch using Adam while the BCFNet with pre-training is optimized by the vanilla SGD. This is because Adam requires momentum information of the previous updated parameters which is not saved in BCFNet with pre-training.

\begin{table}[!t]
	\caption{\black{Statistics of the eight datasets.}}
	\label{statisticsOfDatasets}
	%\centering
	\begin{tabular}{ccccc}
		\hline
		Datasets&\# of Users&\# of Items&\# of Ratings&Sparsity\\
		\hline
		\hline
		ml-100k&943&1682&100000&0.9370\\
		ml-1m&6040&3706&1000209&0.9553\\
		lastfm&1741&2665&69149&0.9851\\
		filmtrust&1508&2071&35497&0.9886\\
		ABaby&746&5193&21262&0.9945\\
		ABeauty&1248&8942&42269&0.9962\\
		AMusic&1776&12929&46087&0.9980 \\
		AToy&3137&33953&84642&0.9992\\
		\hline
	\end{tabular}
\end{table}

\section{Experiments}
\label{sec:experiments}

In this section, we conduct experiments to demonstrate the effectiveness of \black{the \black{BCFNet model}. First of all, we compare the proposed BCFNet model with seven existing models including the previous version namely CFNet\black{~\cite{deng2019deepcf}}. Then, we conduct experiments to validate the effectiveness of the feed-forward attention layer and the balance module. We also verify the utility of pre-training by comparing the BCFNet models with and without pre-training. Finally, we analyze the effect of hyperparameters on the performance of the BCFNet model.} We implement the proposed model based on Keras\footnote{\url{https://github.com/keras-team/keras}} and Tensorflow\footnote{\url{https://github.com/tensorflow/tensorflow}}, which will be released publicly upon acceptance.

\begin{table*}[!t]
	\caption{Comparison results of different methods in terms of HR@10 and NDCG@10. \black{The best scores among the ACFNet model and its sub-models and the best scores among other methods are highlighted respectively in bold.}}
	\label{Comparisons}
	\centering
	
	\begin{tabular}{cc|ccccccc}
		\hline
		\hline
		\multirow{2}*{Datasets}&\multirow{2}*{Measures}&\multicolumn{7}{c}{Existing methods}\\
		& & ItemPop & ItemKNN  & BPR & MLP & DMF & NeuMF & CFNet\\
		\hline
		\multirow{2}*{ml-100k}&HR&0.3998&0.5891&0.6320&0.6755&0.6797&0.6766&\textbf{0.6819}\\
		&NDCG&0.2264&0.3283&0.3568&\textbf{0.3995}&0.3936&0.3945&0.3981\\
		\hline
		\multirow{2}*{ml-1m}&HR&0.4535&0.6624&0.6725&0.7073&0.6565&0.7210&\textbf{0.7253}\\
		&NDCG&0.2542&0.3905&0.3908&0.4264&0.3761&0.4387&\textbf{0.4416}\\
		\hline
		\multirow{2}*{lastfm}&HR&0.6628&0.8771&0.6249&0.8834&0.8840&0.8868&\textbf{0.8995}\\
		&NDCG&0.3862&0.5617&0.3466&0.5919&0.5804&0.6007&\textbf{0.6186}\\
		\hline
		\multirow{2}*{filmtrust}&HR&0.8966&0.8601&0.8680&0.9151&0.9071&\textbf{0.9171}&0.9158\\
		&NDCG&0.7952&0.7582&0.7632&0.8024&0.7896&0.8067&\textbf{0.8074}\\
		\hline
		\multirow{2}*{ABaby}&HR&0.5416&0.2064&0.5751&0.5938&0.5697&\textbf{0.6046}&0.6032\\
		&NDCG&0.3223&0.1170&0.3569&0.3663&0.3479&\textbf{0.3860}&0.3794\\
		\hline
		\multirow{2}*{ABeauty}&HR&0.5938&0.5321&0.6755&0.7099&0.6931&\textbf{0.7260}&0.7123\\
		&NDCG&0.3548&0.3994&0.4738&0.4958&0.4795&\textbf{0.5227}&0.5099\\
		\hline
		\multirow{2}*{AMusic}&HR&0.3148&0.3851&0.3987&0.4071&0.3744&0.3891&\textbf{0.4116}\\
		&NDCG&0.1752&\textbf{0.2825}&0.2420&0.2420&0.2149&0.2391&0.2601\\
		\hline
		\multirow{2}*{AToy}&HR&0.3143&0.3460 &0.3975&0.3931&0.3535&0.3650&\textbf{0.4090}\\
		&NDCG&0.1794&0.2254&\textbf{0.2673}&0.2293&0.2016&0.2155&0.2457\\
		\hline
	\end{tabular}
	\begin{tabular}{cc|cccc|cc}
		\hline
		\multirow{2}*{Datasets}&\multirow{2}*{Measures}&\multicolumn{4}{c|}{ACFNet}&Improvement of&Improvement of\\
		& &  ACFNet-rl & ACFNet-bm & ACFNet-ml & ACFNet & ACFNet vs. NeuMF&ACFNet vs. CFNet\\
		\hline
		\multirow{2}*{ml-100k}&HR&0.6903&0.6681&0.6776&\textbf{0.7010}& 3.61\% & 2.80\% \\
		&NDCG&0.4003&0.3944&0.4011&\textbf{0.4096}& 3.83\% & 2.89\% \\
		\hline
		\multirow{2}*{ml-1m}&HR&0.7199&0.7084&0.7141&\textbf{0.7358}& 2.05\% & 1.45\% \\
		&NDCG&0.4358&0.4342&0.4376&\textbf{0.4496}& 2.48\% & 1.81\% \\
		\hline
		\multirow{2}*{lastfm}&HR&0.8943&0.8897&0.8955&\textbf{0.9110}& 2.73\% & 1.28\% \\
		&NDCG&0.6058&0.6202&0.5970&\textbf{0.6328}& 5.34\% & 2.28\% \\
		\hline
		\multirow{2}*{filmtrust}&HR&0.9151&0.9171&0.9198&\textbf{0.9290}& 1.30\% & 1.44\% \\
		&NDCG&0.8129&0.8099&0.8067&\textbf{0.8231}& 2.03\% & 1.94\% \\
		\hline
		\multirow{2}*{ABaby}&HR&0.6032&0.6059&0.6046&\textbf{0.6086}& 0.66\% & 0.90\% \\
		&NDCG&0.3778&0.3821&0.3708&\textbf{0.3865}& 0.13\% & 1.87\% \\
		\hline
		\multirow{2}*{ABeauty}&HR&0.7179&0.7244&0.7212&\textbf{0.7364}& 1.43\% & 3.38\% \\
		&NDCG&0.5075&0.5272&0.5020&\textbf{0.5299}& 1.38\% & 3.92\% \\
		\hline
		\multirow{2}*{AMusic}&HR&0.4026&0.3958&0.4206&\textbf{0.4448}& 14.32\% & 8.07\% \\
		&NDCG&0.2482&0.2537&0.2492&\textbf{0.2694}& 12.67\% & 3.58\% \\
		\hline
		\multirow{2}*{AToy}&HR&0.3915&0.4080&0.3927&\textbf{0.4201}& 15.10\% & 2.71\% \\
		&NDCG&0.2277&0.2541&0.2270&\textbf{0.2531}& 17.45\% & 3.01\% \\
		\hline
		\hline
	\end{tabular}
	
\end{table*}

\subsection{Experimental Settings}

\subsubsection{Dataset}
We evaluate our models on \black {eight real-world} publicly available datasets: \black{MovieLens 100k (ml-100k)}, MovieLens 1M (ml-1m)\footnote{\url{https://grouplens.org/datasets/movielens/}}, LastFM (lastfm)\footnote{\url{http://www.dtic.upf.edu/~ocelma/MusicRecommendationDataset/}}, \black{FilmTrust (filmtrust)}\footnote{\url{https://www.librec.net/datasets.html}}, \black{Amazon baby (ABaby)}, \black{Amazon beauty (ABeauty)},  Amazon music (AMusic) and Amazon toys (AToy)\footnote{\url{http://jmcauley.ucsd.edu/data/amazon/}}. They are obtained \black{from} the following four main sources. 

\begin{itemize}
\item[1)]
\black{\textbf{MovieLens}: \black{The MovieLens datasets} have been widely used for movie recommendation. These datasets are collected from the MovieLens website by the GroupLens Research. We use the versions ml-100k and ml-1m in our \black{experiments}.}

\item[2)]
\black{
	\textbf{Lastfm}: \black{The} lastfm dataset is a set about the sequence of songs that the users listen to. It is \black{crawled} from the Last.fm online system, which is the world's largest social music platform.}

\item[3)]
\black{\textbf{Filmtrust}: \black{The filmtrust dataset is a dataset crawled} from the entire filmtrust website in June 2011, which contains 1508 users, 2071 items and 35497 ratings.}
\item[4)]
\black{\textbf{Amazon}: \black{The Amazon datasets contain} users' rating data in Amazon. In our experiment, four datasets namely \black{Baby, Beauty, Music and Toy} are adopted.}
	
\end{itemize}

%The ml-1m dataset has been preprocessed by the provider. 
\black{Each dataset has been preprocessed such that each user has at least 20 ratings and each item has been rated by at least 5 users.} The statistics of these eight datasets are summarized in \black{Table}~\ref{statisticsOfDatasets}. \black{Following~\cite{wu2016collaborative,he2017neural}, we construct the user-item interaction matrix for each dataset via Eq.~(\ref{eq:interaction_matrix}).}

\subsubsection{\black{Evaluation Protocols}}
\black{Following~\cite{he2017neural}, we adopt the leave-one-out evaluation, i.e., the latest interaction of each user is used for testing, while the remaining data for training. Since ranking all items is time-consuming, we randomly select 100 unobserved interactions as negative samples for each user. We then rank the 100 items \black{for each user} according to the prediction. We evaluate the model ranking performance through two widely adopted evaluation measures, namely Hit Ratio (HR) and Normalized Discounted Cumulative Gain (NDCG)\black{, which are defined respectively as follows
\begin{align}
HR =& \frac{\#hits}{\#users}\\
NDCG =& \frac{1}{\#users}\sum_{i=1}^{\#users}\frac{1}{\log_{2}(p_{i}+1)}
\end{align} 
where} $\#hits$ is the number of users whose test item appears in the recommended list and $p_i$ is the position of the test item in the list for the $i$-th hit. The ranked list is truncated at 10 for both measures\black{, i.e. HR@10 and NDCG@10. Intuitively, HR@10 measures whether the test item is present on the top-10 list or not, and NDCG@10 measures the ranking quality which assigns higher scores to hit at top position ranks on the top-10 list. Larger values of HR@10 and NDCG@10 indicate the better performance.}}

\subsubsection{\black{Parameter Settings}}
\black{For a fair comparison, we set the weight of observed interactions for each user as 1 for all methods. We sample four negative instances per positive instance, i.e., set the negative sample ratio $\rho$ to be 4 as default. We set the number of predictive factors as 128 on all the datasets except AMusic, on which the number of predictive factor is set as 64. We generally employ two hidden layers for \black{BCFNet-rl}, and three for \black{BCFNet-ml}.}

\subsection{Comparison Results}
\black{We compare the proposed \black{BCFNet model} with the following seven methods.}
\black{
\begin{itemize}
	\item
	\textbf{ItemPop} is a non-personalized method that is often used as a benchmark for recommendation tasks. Items are simply ranked by their popularity, i.e., the number of interactions.
	\item
	\textbf{ItemKNN}~\cite{sarwar2001item} is a standard item-based collaborative filtering method. 
	\item
	\textbf{BPR}~\cite{rendle2009bpr} is a widely used learning framework for item recommendation with implicit feedback. It is a sample-based method that optimizes the MF model with a \black{pair-wise} ranking loss.
	\item
	\textbf{MLP}~\cite{he2017neural} is a matching \black{function} leaning-based collaborative filtering method. It uses multiple layers of nonlinearities to model the relationships between users and items.
	\item
	\textbf{DMF}~\cite{xue2017deep} is a state-of-the-art representation learning-based MF method which performs deep matrix factorization to learn a common low dimensional space with normalized cross entropy loss as loss function. It uses a two-pathway neural network architecture to replace the linear embedding operation used in vanilla matrix factorization. We ignore the explicit ratings and take the implicit feedback as input in this paper.
	\item
	\textbf{NeuMF}~\cite{he2017neural} is a state-of-the-art matching function learning-based MF method which combines hidden layers of GMF and MLP to learn the interaction function based on cross entropy loss. NeuMF takes IDs as input and adapts the deep+shallow pattern which has been widely adopted in many works such as~\cite{cheng2016wide,guo2017deepfm}.
	\item
	\black{\textbf{CFNet}~\cite{deng2019deepcf} is the previous version of BCFNet, which incorporates collaborative filtering methods based on representation learning and matching function learning to learn the complex matching function and low-rank relations between users and items.}
\end{itemize}}

The comparison results are listed in \black{Table}~\ref{Comparisons}. \black{The best scores among the BCFNet model and its sub-models and the best scores among other methods are \black{highlighted} respectively in bold.} According to the table, we have the following key observations:
\begin{itemize}
	\item \black{The proposed BCFNet model \black{achieves the best performance on} all the datasets \black{except for the NDCG on AMusic and Atoy,} and obtains high improvements over the state-of-the-art methods. More importantly, most of the improvements increase} along with the increasing of data sparsity, where the datasets are arranged in the order of increasing data sparsity. This justifies the effectiveness of the proposed \black{BCFNet model} \black{which combines attentive representation learning-based CF methods, attentive matching function learning-based CF methods and balance module.}
	\item \black{As a typical representation learning method, the performance of DMF has some merit compared with the traditional methods, but the proposed BCFNet-rl model} consistently outperforms it. \black{This indicates that adding a feed-forward attention layer and a parametric neural network layer significantly \black{improves} the learning ability of the representation leaning.}
	\item \black{Compared with \black{the} MLP model, BCFNet-ml outperforms it on most datasets. This fully demonstrates the effectiveness of attention mechanism in improving matching function learning.}
	\item \black{On the basis of BCFNet-rl and BCFNet-ml, BCFNet-bm has also made a great contribution to the effects of the proposed BCFNet model, especially the improvement \black{in terms of} NDCG. And in most cases, the \black{improvement} effect increases with the increase of data sparsity. \black{This indicates the effectiveness of the balance module in addressing the overfitting issue caused by data sparsity, which will be further confirmed in the next subsection.}}
\end{itemize}

\begin{figure}[!t]
	\centering
	\includegraphics[width=1.0\linewidth]{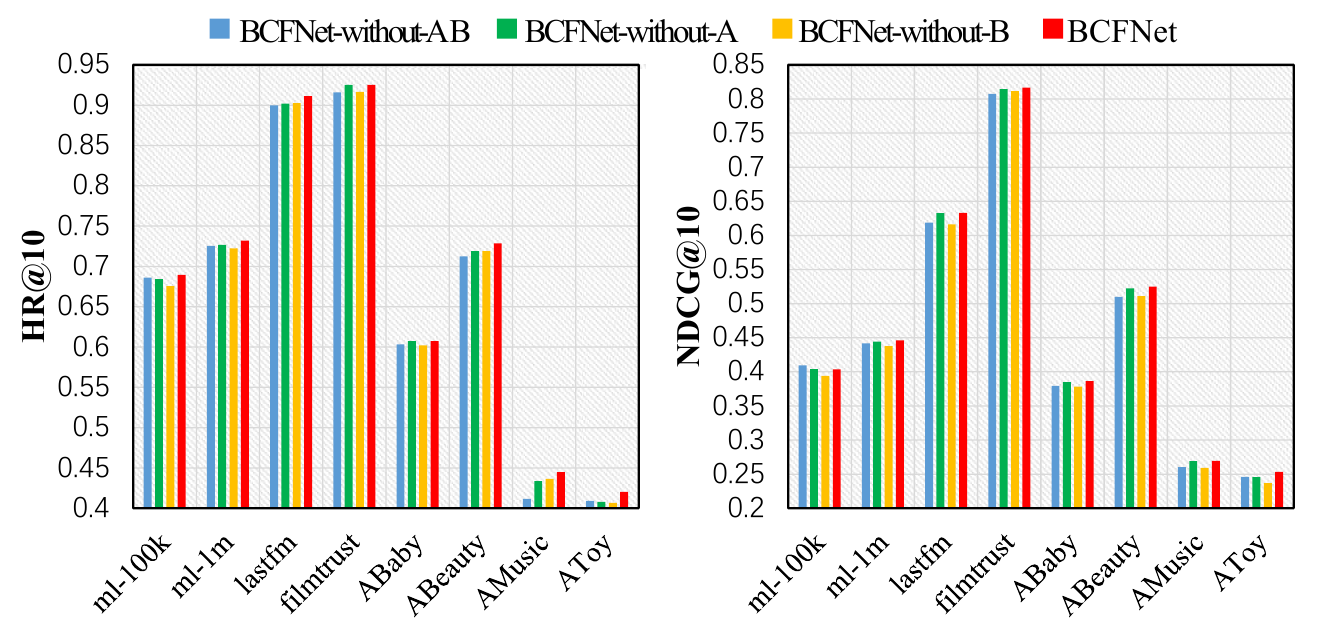}
	\caption{The impact of feed attention layer and balance module on performance on the eight datasets.}
	\label{fig:attention_and_balance}
\end{figure}

	\begin{figure*}[!t]
	\centerline{
	\subfigure[lastfm]{
		\includegraphics[width=0.333\linewidth]{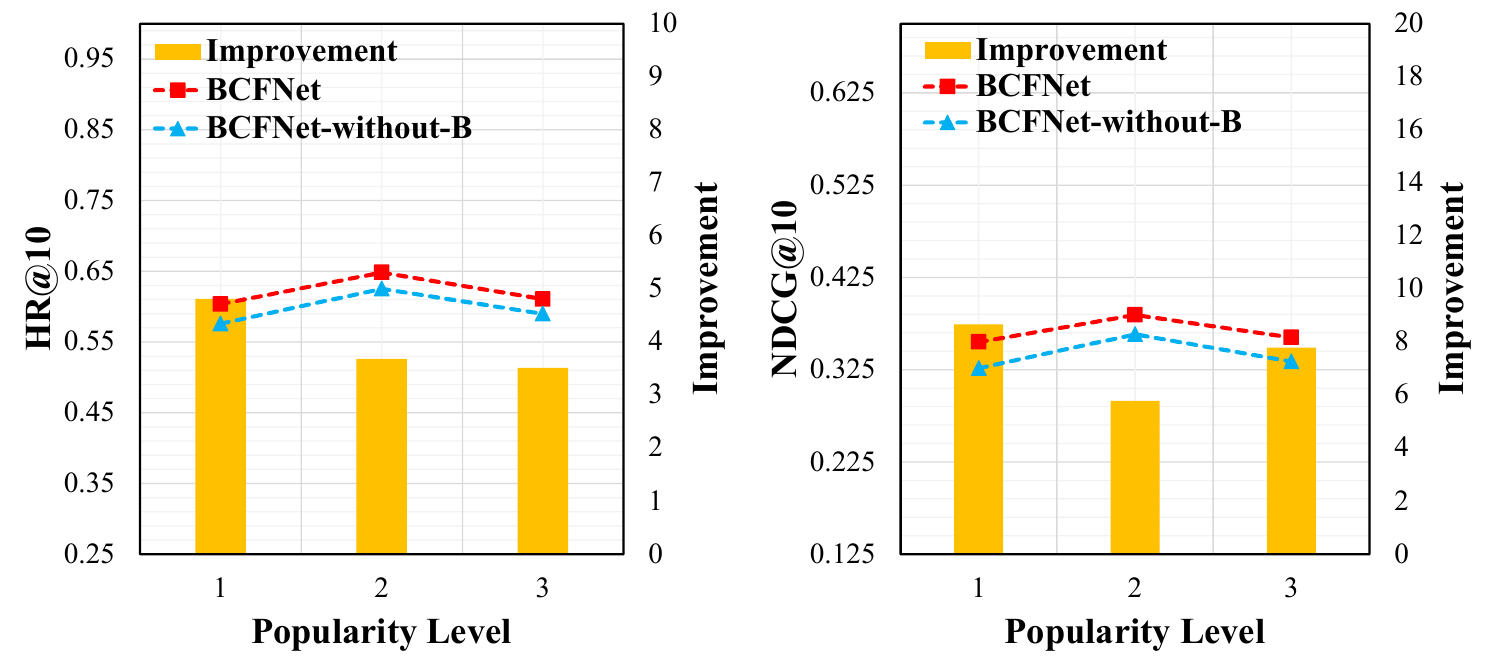}
	}
	\subfigure[ABaby]{
		\includegraphics[width=0.333\linewidth]{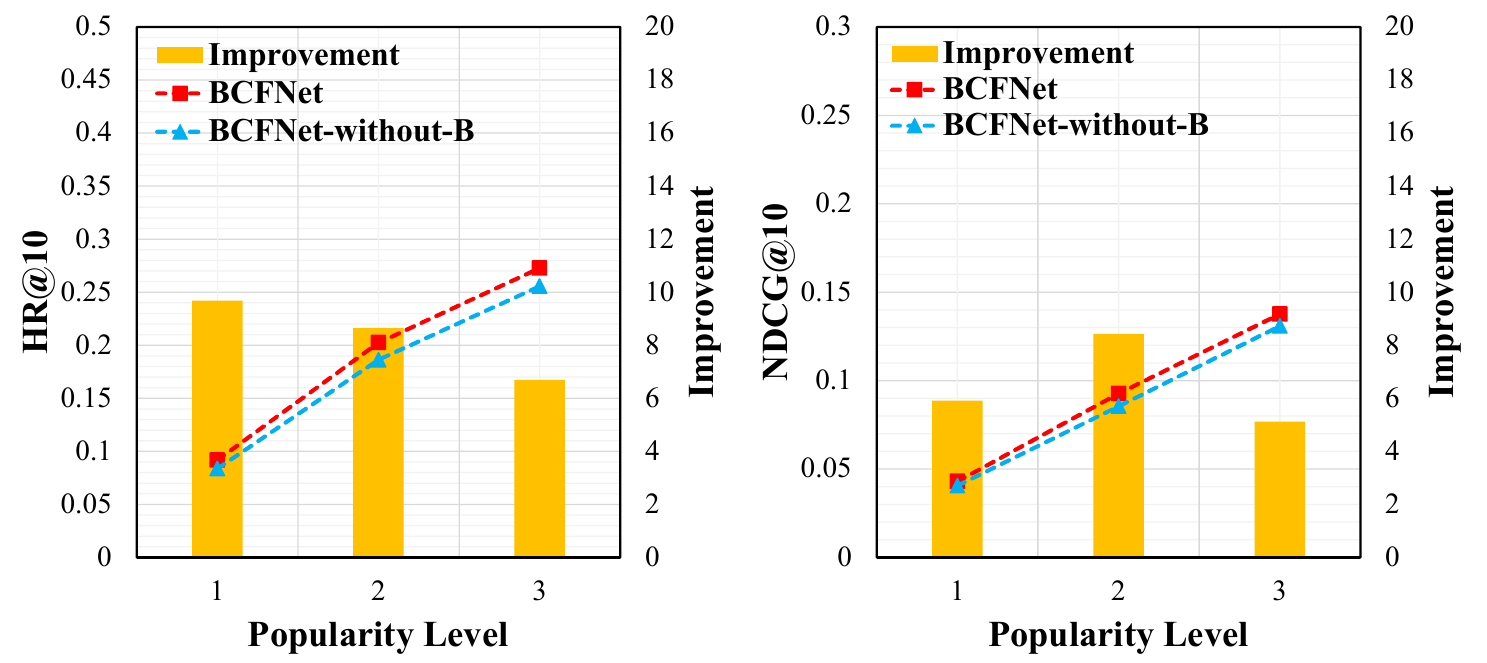}
		}
	\subfigure[ABeauty]{
		\includegraphics[width=0.333\linewidth]{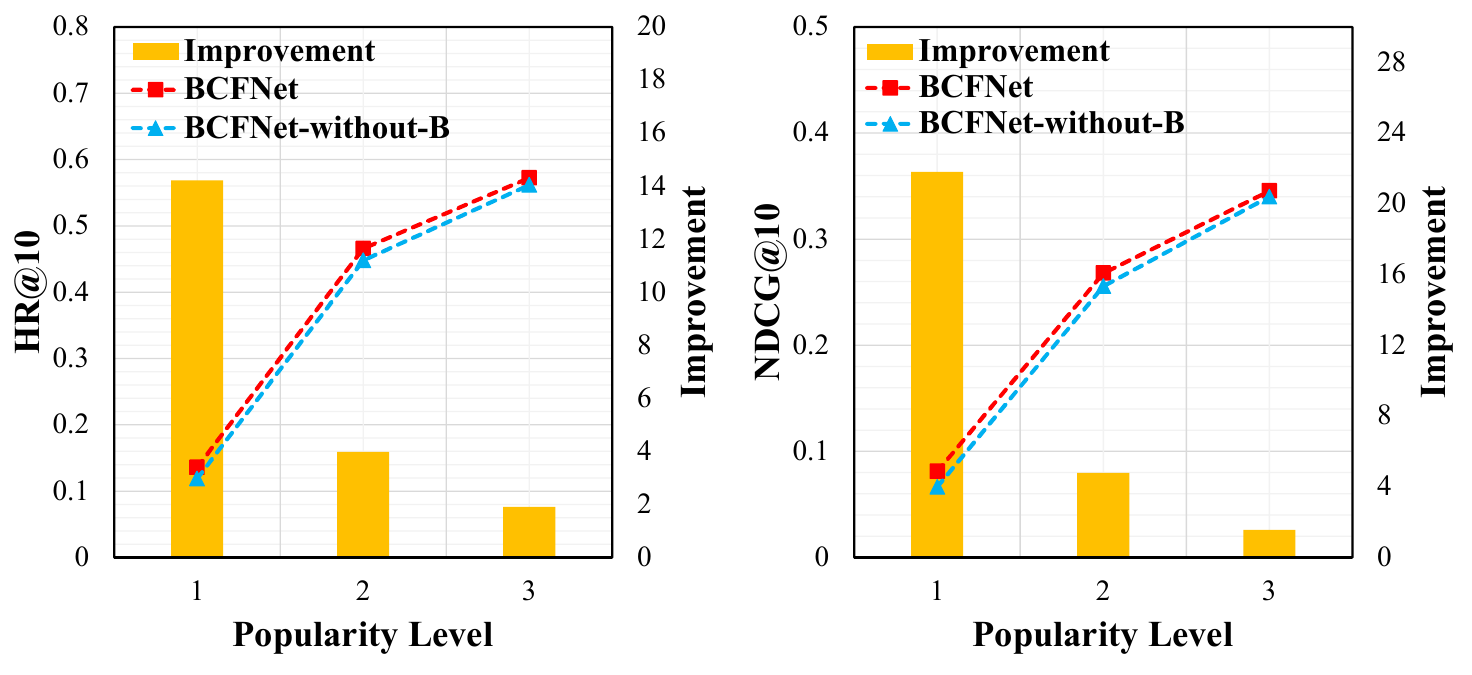}
	}
	}
		\centerline{
	\subfigure[AMusic]{
		\includegraphics[width=0.333\linewidth]{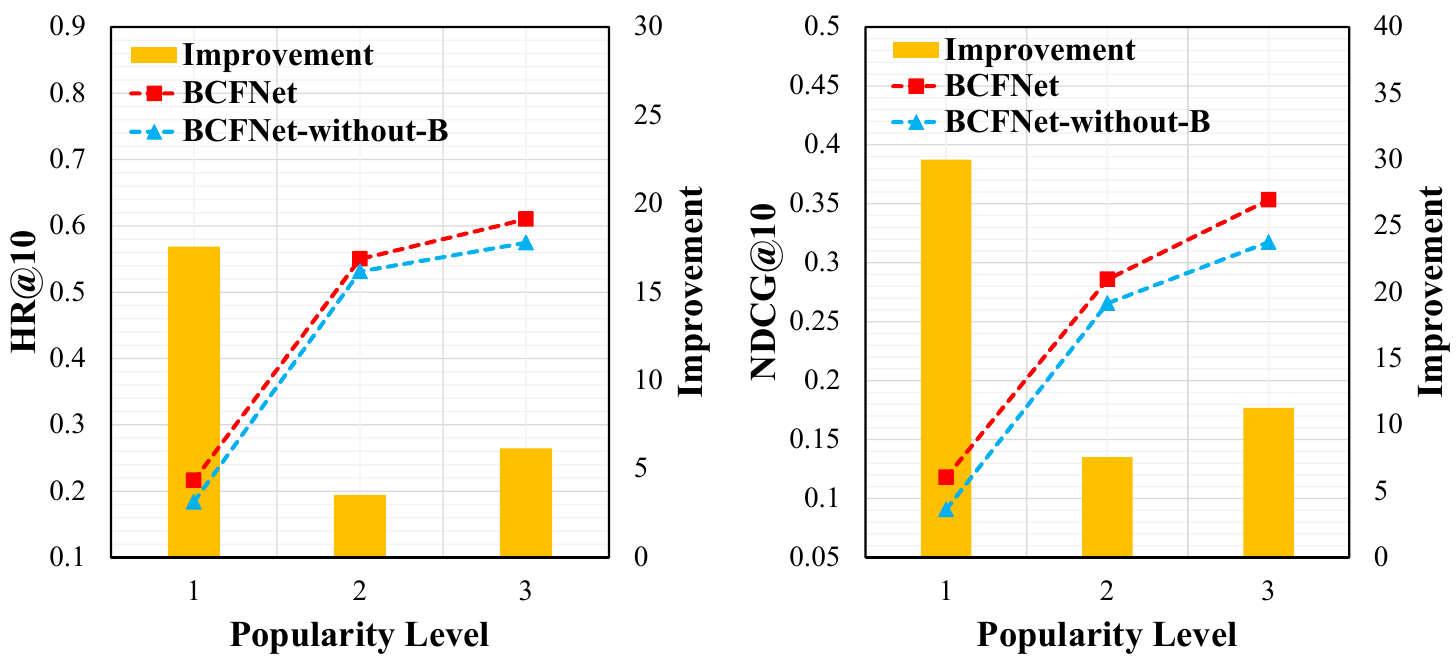}
	}
	\subfigure[AToy]{
		\includegraphics[width=0.333\linewidth]{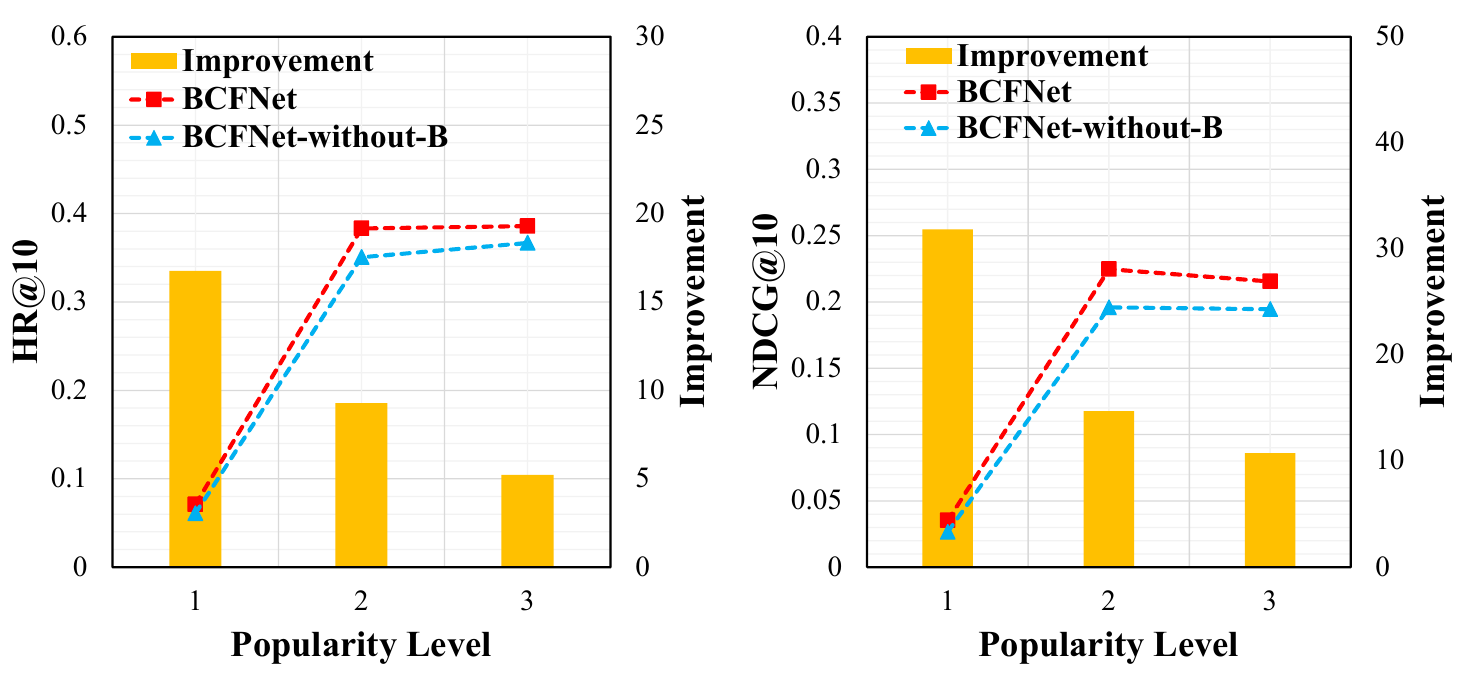}
		}
	}
	%\vskip -0.2in
	\caption{\black{The impact of balance module on the performance obtained on sub-datasets with different item popularity levels \black{on the five datasets}.}}
	\label{fig:balance_overfitting}
\end{figure*}

	\subsection{\black{Impact of the Feed-forward Attention Layer and the Balance Module}}
	\black{In order to investigate the impact of feed-forward attention layer and balance module in BCFNet, we conduct experiments on BCFNet without attention and balance module (\textit{abbr.} BCFNet-without-AB, i.e. CFNet in~\cite{deng2019deepcf}), BCFNet without attention (\textit{abbr.} BCFNet-without-A) and BCFNet without balance module (\textit{abbr.} BCFNet-without-B). As shown in \figurename~\ref{fig:attention_and_balance}, BCFNet outperforms BCFNet-without-AB, BCFNet-without-A and BCFNet-without-B in all cases. This result verifies the effectiveness of the feed-forward attention layer in enhancing the learning ability of the proposed neural network model. Moreover, BCFNet-without-AB outperforms BCFNet-without-A and BCFNet-without-B on some datasets, which shows the necessity of combining the feed-forward attention layer and the balance module.}

\black{In addition, we also conduct more experiments on BCFNet with balance module (i.e. the BCFNet model) and BCFNet without balance module (\textit{abbr.} BCFNet-without-B) to verify the effectiveness of the balance module in alleviating over-fitting issue of neural network. In order to simulate the over-fitting issue caused by the high sparsity of item interaction information in a recommender system, we divide \black{some original dataset into three sub-datasets according to item popularity, which are termed popularity \black{levels} 1, 2 and 3 respectively. \black{A higher popularity level means that items in this sub-dataset are more popular and have more interaction  information.} In particular, for some original dataset, the item set is evenly partitioned into three subsets according to item popularity, and then all the interactions associated with items in each subset form a corresponding sub-dataset. Therefore, the sparsity of interaction information will decrease with the increase of item popularity level. Since some datasets used in experiments are not very sparse or cannot satisfy the leave-one-out evaluation condition that requires 100 negative samples for each user in their three sub-datasets, so only five datasets are used in this experiment, namely lastfm, ABaby, ABeauty, AMusic and AToy.}
\black {We} run BCFNet and BCFNet-without-B respectively on each sub-dataset. As shown in \figurename~\ref{fig:balance_overfitting}, with the increase of item popularity level, the effect of BCFNet and BCFNet-without-B will be greatly improved. However, with the increase of item popularity level, most of the promotion effect of the balance module will be weakened, i.e. the highest promotion effect has been obtained in the case of the smallest item popularity. This fully illustrates that the balance module is helpful to alleviate the over-fitting issue caused by the high sparsity of item interaction information.
}

\begin{figure*}[!t]
	\centerline{
		\subfigure[ml-100k]{
			\includegraphics[width=0.333\linewidth]{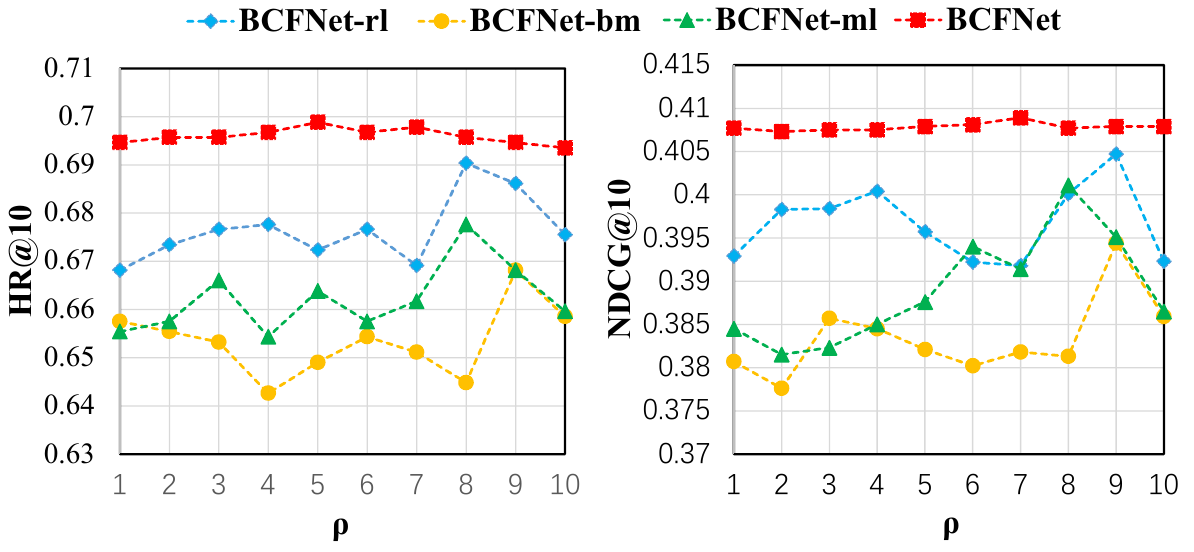}
		}
		\subfigure[ml-1m]{
			\includegraphics[width=0.333\linewidth]{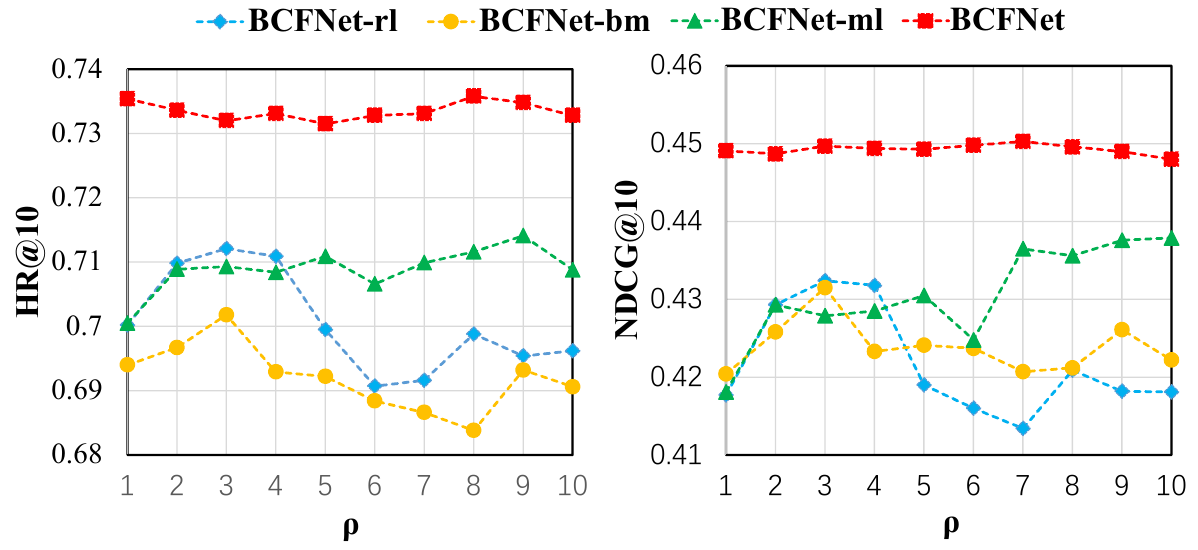}
		}
		\subfigure[lastfm]{
			\includegraphics[width=0.333\linewidth]{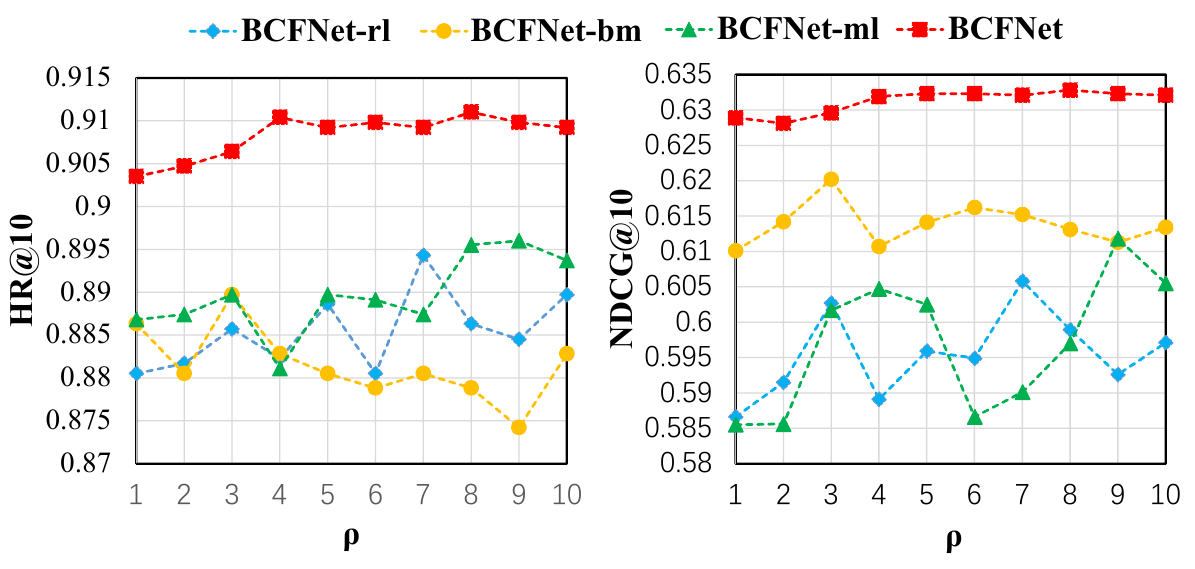}
		}
	}
	\centerline{
		\subfigure[filmtrust]{
			\includegraphics[width=0.333\linewidth]{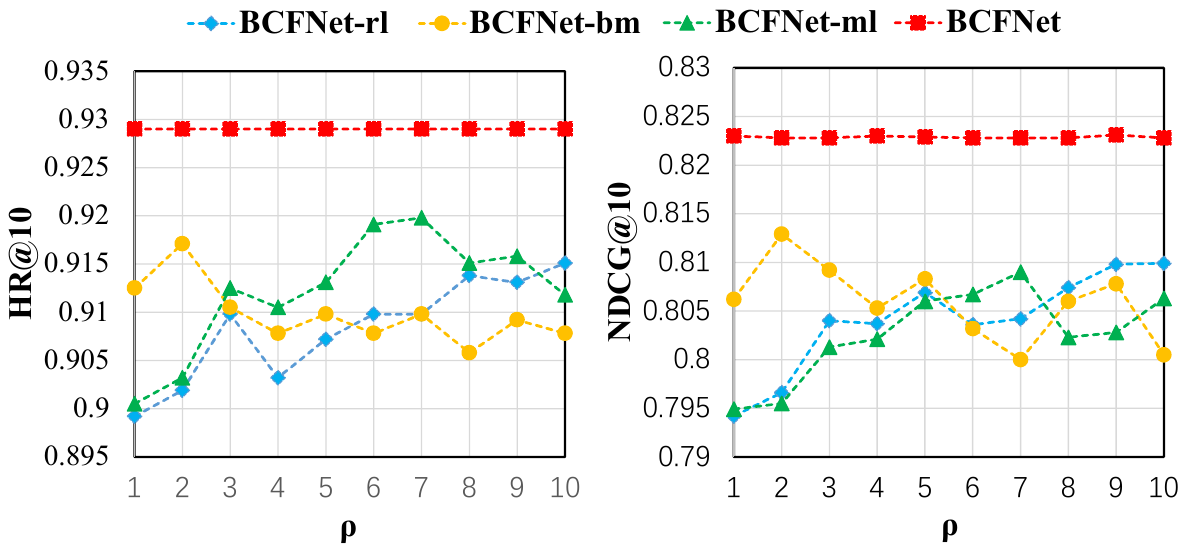}
		}
		\subfigure[ABaby]{
			\includegraphics[width=0.333\linewidth]{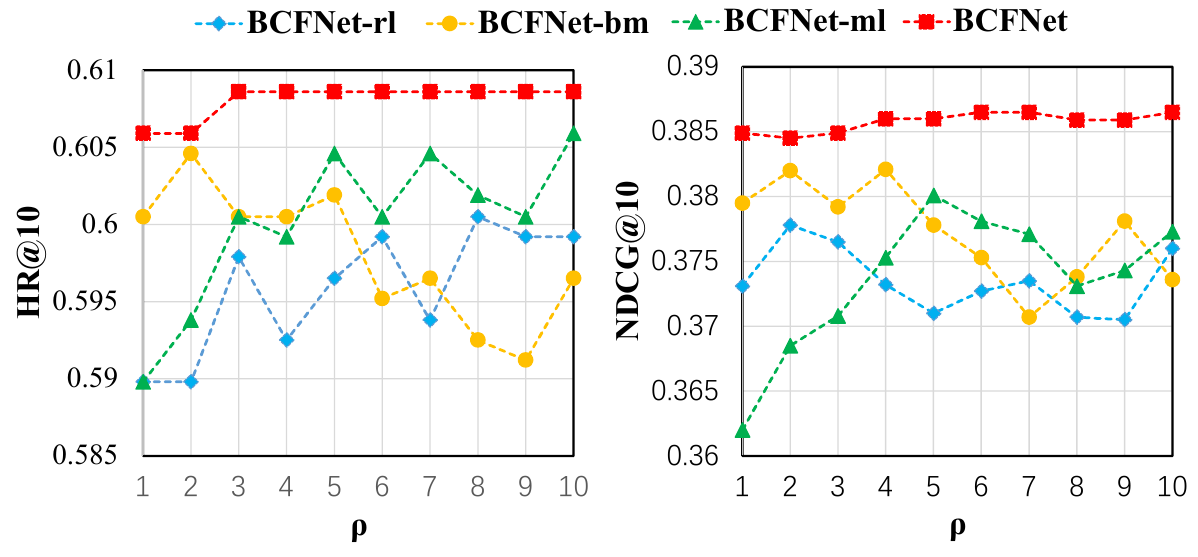}
		}
		\subfigure[ABeauty]{
			\includegraphics[width=0.333\linewidth]{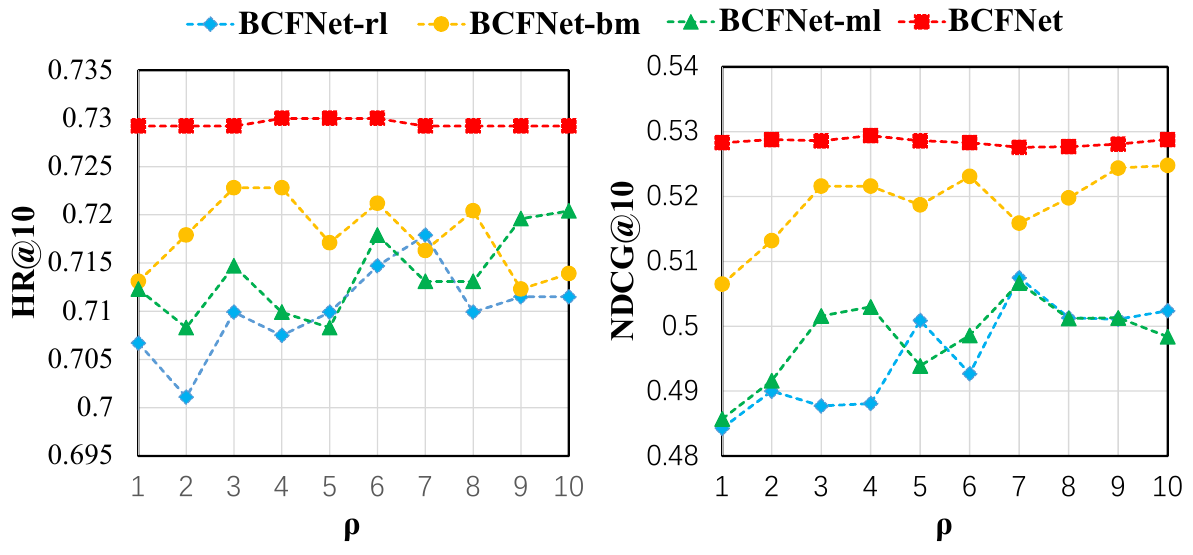}
		}
	}
	\centerline{
		\subfigure[AMusic]{
			\includegraphics[width=0.333\linewidth]{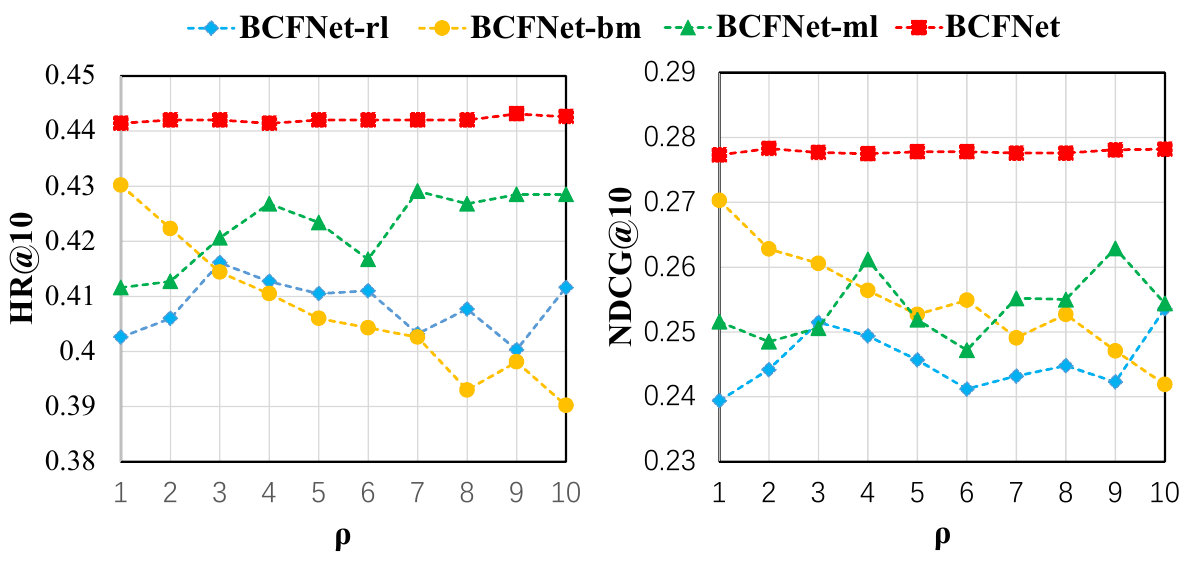}
		}
		\subfigure[AToy]{
			\includegraphics[width=0.333\linewidth]{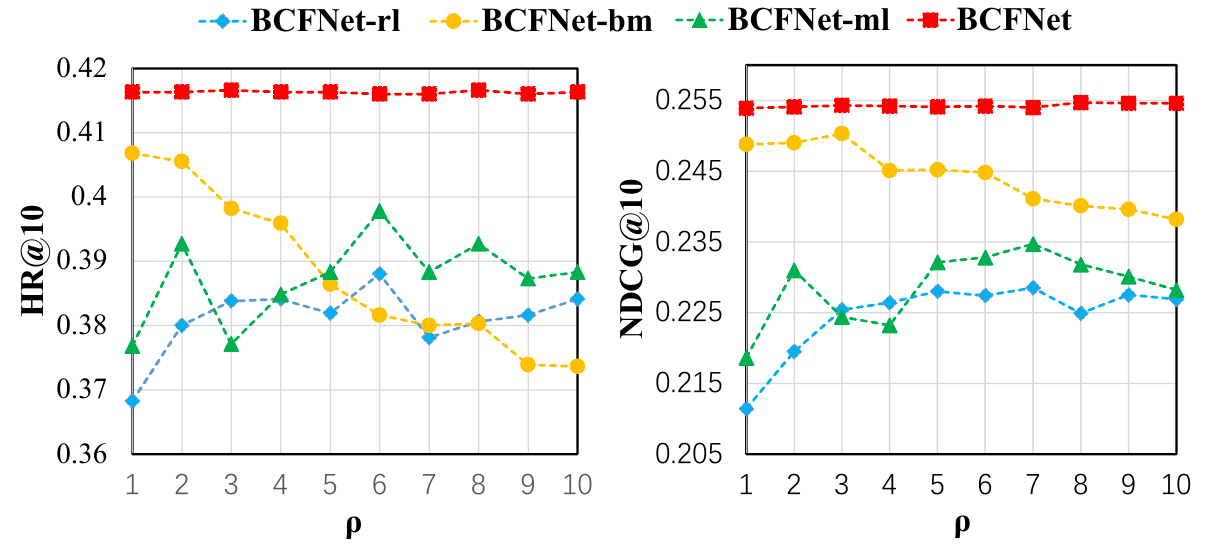}
		}
	}
	%\vskip -0.2in
	\caption{\black{The effect of negative sampling ratio \black{$\rho$} on the performance of BCFNet and its three sub-models.}}
	\label{fig:parameter_sensitivity_neg}
\end{figure*}

\subsection{Impact of Pre-training}
Different from the BCFNet with pre-training, we use mini-batch Adam to learn the BCFNet without pre-training with random initializations. As shown in \black{Table}~\ref{Pretraining}, \black{the BCFNet with pre-training (i.e. BCFNet) outperforms the BCFNet without pre-training (\textit{abbr.} BCFNet-without-P) in all cases.} This result verifies the utility of the pre-training process which ensures BCFNet-rl, BCFNet-ml and BCFNet-bm to learn features from different perspectives and therefore allows the model to generate better results.

\subsection{Sensitivity Analysis of Hyperparameters}

\subsubsection{Negative Sampling Ratio} 
To analyze the effect of negative sampling ratio \black{$\rho$}, we test different negative sampling ratio, i.e., the number of negative samples per positive instance, on the \black{eight datasets}. From the results shown in \figurename~\ref{fig:parameter_sensitivity_neg}, we can find that sampling \black{less than three instances} is not enough and sampling more negative instances is helpful. \black{\black{In most cases}, the best HR@10 and NDCG@10 are obtained when the negative sampling ratio is \black{set to 4}}. Overall, the optimal sampling ratio is around \black{4 to 8}. Sampling more negative instances not only requires more time to train the model but also degrades the performance, which is consistent with the results shown in~\cite{he2017neural}.

\subsubsection{The Number of Predictive Factors} 

Another hyperparameter used in the BCFNet model is the number of predictive factors, i.e., \black{the dimensions of $\mathbf{a}_\mathrm{Y}^{rl}$, $\mathbf{a}_\mathrm{Y}^{ml}$ and $\mathbf{a}_\mathrm{Y}^{bm}$.} \black{To this end, we test the number of predictive factors in \black{$[16, 32, 64, 128]$}, and the results are listed in Table~\ref{Factors}. As} shown in \black{Table}~\ref{Factors}, the proposed model generates the best performance with \black{128} predictive factors on most of the datasets except the AMusic dataset. On the AMusic dataset, the best performance is achieved with \black{64} factors. According to our observation, more predictive factors usually lead to better performances since it endows the model with larger capability and greater ability of representation.

\begin{table}[!t]
	\caption{Performance of BCFNet without pre-training (\black{BCFNet-without-P}) and BCFNet with pre-training (BCFNet).}
	\label{Pretraining}
	\centering
	\begin{tabular}{@{}c@{}c@{}|@{}c@{}c@{}|@{}c@{}}
		\hline
		\hline
		~~Datasets~~&~~Measures~~&~~\black{BCFNet-without-P}~~&~~\black{BCFNet}~~&~~Improvement\\
		\hline
		\multirow{2}*{ml-100k}&HR&0.5769 &0.7010& 21.51\%  \\
		&NDCG&0.3216&0.4096& 27.36\%  \\
		\hline
		\multirow{2}*{ml-1m}&HR&0.6843&0.7358& 7.53\%  \\
		&NDCG&0.4099&0.4496& 9.69\%  \\
		\hline
		\multirow{2}*{lastfm}&HR&0.8621&0.9110& 5.67\%  \\
		&NDCG&0.5871&0.6328& 7.78\%  \\
		\hline
		\multirow{2}*{filmtrust}&HR&0.8899&0.9290& 4.39\%  \\
		&NDCG&0.7903&0.8231& 4.15\%  \\
		\hline
		\multirow{2}*{ABaby}&HR&0.5456&0.6086& 11.55\%  \\
		&NDCG&0.3309&0.3865& 16.80\%  \\
		\hline
		\multirow{2}*{ABeauty}&HR&0.7099&0.7364& 3.73\%  \\
		&NDCG&0.4805&0.5299& 10.28\%  \\
		\hline
		\multirow{2}*{AMusic}&HR&0.3874&0.4448& 14.82\%  \\
		&NDCG&0.2356&0.2694& 14.35\%  \\
		\hline
		\multirow{2}*{AToy}&HR&0.3028&0.4201& 38.74\%  \\
		&NDCG&0.1631&0.2531& 55.18\%  \\
		\hline
		\hline
	\end{tabular}
\end{table}

\section{Conclusion and Future Work}
\label{sec:conclusion}
\black{In this paper, we have presented a \black{novel recommendation model called Balanced Collaborative Filtering Network (BCFNet)}, which combines attentive representation learning (BCFNet-rl), attentive matching function learning (BCFNet-ml) and balance module (BCFNet-bm). Therefore, it has the advantages of both representation learning and matching function learning. In addition, by introducing a feed-forward attention layer, the learning ability of both of attentive representation learning and attentive matching function learning can be further improved. Furthermore, adding a balance module without using neural network and attention mechanism can alleviate the over-fitting issue and capture low-rank relation. Extensive experiments on eight real-world datasets demonstrate the effectiveness and rationality of the proposed BCFNet model.}

\section{Acknowledgments}
This work was supported by NSFC (61876193), Guangdong Natural Science Funds for Distinguished Young Scholar (2016A030306014), and NSF through grants IIS-1526499, IIS-1763325, and CNS-1626432.

\begin{table}[!t]
	\caption{Performance of BCFNet with different number of predictive factors.}
	\label{Factors}
	\centering
	\begin{tabular}{@{}c@{}c@{}|cccc@{}}
		\hline
		\hline
		\multirow{2}*{Datasets}&\multirow{2}*{~~Measures~~}&\multicolumn{4}{c}{Dimensions of predictive vectors}\\
		&&16&32&64&128\\
		\hline
		\multirow{2}*{ml-100k}&HR&0.6660&0.6723&0.6702&0.7010\\
		&NDCG&0.3850&0.3896&0.3912&0.4096\\
		\hline
		\multirow{2}*{ml-1m}&HR&0.6980&0.7078&0.7230&0.7358\\
		&NDCG&0.4162&0.4261&0.4396&0.4496\\
		\hline
		\multirow{2}*{lastfm}&HR&0.8926&0.8909&0.9012&0.9110\\
		&NDCG&0.6219&0.6231&0.6250&0.6328\\
		\hline
		\multirow{2}*{filmtrust}&HR&0.9045&0.9131&0.9204&0.9290\\
		&NDCG&0.7947&0.8032&0.8101&0.8231\\
		\hline
		\multirow{2}*{ABaby}&HR&0.6005&0.5965&0.6072&0.6086\\
		&NDCG&0.3765&0.3800&0.3855&0.3865\\
		\hline
		\multirow{2}*{ABeauty}&HR&0.7163&0.7171&0.7212&0.7364\\
		&NDCG&0.5020&0.5056&0.5133&0.5299\\
		\hline
		\multirow{2}*{AMusic}&HR&0.4110&0.4245&0.4448&0.4240\\
		&NDCG&0.2581&0.2608&0.2694&0.2638\\
		
		\hline
		\multirow{2}*{AToy}&HR&0.4080&0.4013&0.4074&0.4201\\
		&NDCG&0.2464&0.2418&0.2495&0.2531\\
		
		\hline
		\hline
	\end{tabular}
\end{table}

% % % % % % % % % %
% References and End of Paper

\bibliography{ref}

% Generated by IEEEtran.bst, version: 1.13 (2008/09/30)
\begin{thebibliography}{10}
\providecommand{\url}[1]{#1}
\csname url@samestyle\endcsname
\providecommand{\newblock}{\relax}
\providecommand{\bibinfo}[2]{#2}
\providecommand{\BIBentrySTDinterwordspacing}{\spaceskip=0pt\relax}
\providecommand{\BIBentryALTinterwordstretchfactor}{4}
\providecommand{\BIBentryALTinterwordspacing}{\spaceskip=\fontdimen2\font plus
\BIBentryALTinterwordstretchfactor\fontdimen3\font minus
  \fontdimen4\font\relax}
\providecommand{\BIBforeignlanguage}[2]{{%
\expandafter\ifx\csname l@#1\endcsname\relax
\typeout{** WARNING: IEEEtran.bst: No hyphenation pattern has been}%
\typeout{** loaded for the language `#1'. Using the pattern for}%
\typeout{** the default language instead.}%
\else
\language=\csname l@#1\endcsname
\fi
#2}}
\providecommand{\BIBdecl}{\relax}
\BIBdecl

\bibitem{Wang_Serendipitous:18}
C.-D. Wang, Z.-H. Deng, J.-H. Lai, and P.~S. Yu, ``Serendipitous recommendation
  in {E}-commerce using innovator-based collaborative filtering,'' \emph{{IEEE}
  Trans. Cybernetics}, vol.~49, no.~7, pp. 2678--2692, 2019.

\bibitem{Zhao_LSCD:18}
Z.-L. Zhao, L.~Huang, C.-D. Wang, and D.~Huang, ``Low-rank and sparse
  cross-domain recommendation algorithm,'' in \emph{DASFAA}, 2018, pp.
  150--157.

\bibitem{Hu_ItemRec:17}
Q.-Y. Hu, Z.-L. Zhao, C.-D. Wang, and J.-H. Lai, ``An item orientated
  recommendation algorithm from the multi-view perspective,''
  \emph{Neurocomputing}, vol. 269, pp. 261--272, 2017.

\bibitem{Cai2014}
Y.~Cai, H.~fung Leung, Q.~Li, H.~Min, J.~Tang, and J.~Li, ``Typicality-based
  collaborative filtering recommendation,'' \emph{{IEEE} Trans. Knowl. Data
  Eng.}, vol.~26, no.~3, pp. 766--779, 2014.

\bibitem{Wang2015rctr}
H.~{Wang} and W.~{Li}, ``Relational collaborative topic regression for
  recommender systems,'' \emph{{IEEE} Trans. Knowl. Data Eng.}, pp. 1343--1355,
  2015.

\bibitem{Hu_Item:18}
Q.-Y. Hu, L.~Huang, C.-D. Wang, and H.-Y. Chao, ``Item orientated
  recommendation by multi-view intact space learning with overlapping,''
  \emph{Knowl. Based Syst.}, vol. 164, pp. 358--370, 2019.

\bibitem{he2016deep}
K.~He, X.~Zhang, S.~Ren, and J.~Sun, ``Deep residual learning for image
  recognition,'' in \emph{CVPR}, 2016, pp. 770--778.

\bibitem{graves2013speech}
A.~Graves, A.-r. Mohamed, and G.~Hinton, ``Speech recognition with deep
  recurrent neural networks,'' in \emph{ICASSP}, 2013, pp. 6645--6649.

\bibitem{serban2016building}
I.~V. Serban, A.~Sordoni, Y.~Bengio, A.~C. Courville, and J.~Pineau, ``Building
  end-to-end dialogue systems using generative hierarchical neural network
  models,'' in \emph{AAAI}, 2016, pp. 3776--3784.

\bibitem{Bai2019nlp}
X.~{Bai}, R.~{Abasi}, B.~{Edizel}, and A.~{Mantrach}, ``Position-aware deep
  character-level {CTR} prediction for sponsored search,'' \emph{{IEEE} Trans.
  Knowl. Data Eng.}, 2019.

\bibitem{Huang2013dssm}
P.-S. Huang, X.~He, J.~Gao, L.~Deng, A.~Acero, and L.~P. Heck, ``Learning deep
  structured semantic models for web search using clickthrough data,'' in
  \emph{{CIKM}}, 2013, pp. 2333--2338.

\bibitem{xue2017deep}
H.-J. Xue, X.-Y. Dai, J.~Zhang, S.~Huang, and J.~Chen, ``Deep matrix
  factorization models for recommender systems,'' in \emph{IJCAI}, 2017, pp.
  3203--3209.

\bibitem{xu2018deep}
J.~Xu, X.~He, and H.~Li, ``Deep learning for matching in search and
  recommendation,'' in \emph{SIGIR Tutorial}, 2018, pp. 1365--1368.

\bibitem{hornik1989multilayer}
K.~Hornik, M.~Stinchcombe, and H.~White, ``Multilayer feedforward networks are
  universal approximators,'' \emph{Neural networks}, vol.~2, no.~5, pp.
  359--366, 1989.

\bibitem{he2017neural}
X.~He, L.~Liao, H.~Zhang, L.~Nie, X.~Hu, and T.-S. Chua, ``Neural collaborative
  filtering,'' in \emph{WWW}, 2017, pp. 173--182.

\bibitem{beutel2018latent}
A.~Beutel, P.~Covington, S.~Jain, C.~Xu, J.~Li, V.~Gatto, and E.~H. Chi,
  ``Latent cross: Making use of context in recurrent recommender systems,'' in
  \emph{WSDM}, 2018, pp. 46--54.

\bibitem{deng2019deepcf}
Z.-H. Deng, L.~Huang, C.-D. Wang, J.-H. Lai, and P.~S. Yu, ``{DeepCF}: {A}
  unified framework of representation learning and matching function learning
  in recommender system,'' in \emph{{AAAI}}, 2019, pp. 61--68.

\bibitem{oard1998implicit}
D.~W. Oard, J.~Kim \emph{et~al.}, ``Implicit feedback for recommender
  systems,'' in \emph{Proceedings of the AAAI workshop on recommender systems},
  vol.~83, 1998.

\bibitem{ma2013experimental}
H.~Ma, ``An experimental study on implicit social recommendation,'' in
  \emph{SIGIR}, 2013, pp. 73--82.

\bibitem{hu2008collaborative}
Y.~Hu, Y.~Koren, and C.~Volinsky, ``Collaborative filtering for implicit
  feedback datasets,'' in \emph{ICDM}, 2008, pp. 263--272.

\bibitem{koren2008factorization}
Y.~Koren, ``Factorization meets the neighborhood: a multifaceted collaborative
  filtering model,'' in \emph{KDD}, 2008, pp. 426--434.

\bibitem{rendle2009bpr}
S.~Rendle, C.~Freudenthaler, Z.~Gantner, and L.~Schmidt-Thieme, ``{BPR:
  Bayesian} personalized ranking from implicit feedback,'' in \emph{UAI}, 2009,
  pp. 452--461.

\bibitem{mnih2012learning}
A.~Mnih and Y.~W. Teh, ``Learning label trees for probabilistic modelling of
  implicit feedback,'' in \emph{NIPS}, 2012, pp. 2816--2824.

\bibitem{he2016vbpr}
R.~He and J.~McAuley, ``{VBPR}: Visual bayesian personalized ranking from
  implicit feedback,'' in \emph{AAAI}, 2016, pp. 144--150.

\bibitem{funk2006svd}
S.~Funk, ``Netflix update: Try this at home,''
  \url{http://sifter.org/~simon/journal/20061211.html}, 2006, online; accessed
  27-June-2018.

\bibitem{salakhutdinov2008bayesian}
R.~Salakhutdinov and A.~Mnih, ``Bayesian probabilistic matrix factorization
  using {Markov chain Monte Carlo},'' in \emph{ICML}, 2008, pp. 880--887.

\bibitem{koren2009matrix}
Y.~Koren, R.~Bell, and C.~Volinsky, ``Matrix factorization techniques for
  recommender systems,'' \emph{Computer}, vol.~42, no.~8, 2009.

\bibitem{koren2009collaborative}
Y.~Koren, ``Collaborative filtering with temporal dynamics,'' in \emph{KDD},
  2009, pp. 447--456.

\bibitem{hu2014your}
L.~Hu, A.~Sun, and Y.~Liu, ``Your neighbors affect your ratings: on
  geographical neighborhood influence to rating prediction,'' in \emph{SIGIR},
  2014, pp. 345--354.

\bibitem{Wang2015cdl}
H.~Wang, N.~Wang, and D.-Y. Yeung, ``Collaborative deep learning for
  recommender systems,'' in \emph{KDD}, 2015, pp. 1235--1244.

\bibitem{sedhain2015autorec}
S.~Sedhain, A.~K. Menon, S.~Sanner, and L.~Xie, ``{AutoRec}: Autoencoders meet
  collaborative filtering,'' in \emph{WWW}, 2015, pp. 111--112.

\bibitem{bai2017neural}
T.~Bai, J.-R. Wen, J.~Zhang, and W.~X. Zhao, ``A neural collaborative filtering
  model with interaction-based neighborhood,'' in \emph{CIKM}, 2017, pp.
  1979--1982.

\bibitem{cheng2016wide}
H.-T. Cheng, L.~Koc, J.~Harmsen, T.~Shaked, T.~Chandra, H.~Aradhye,
  G.~Anderson, G.~Corrado, W.~Chai, M.~Ispir \emph{et~al.}, ``Wide \& deep
  learning for recommender systems,'' in \emph{Proceedings of the 1st Workshop
  on Deep Learning for Recommender Systems}, 2016, pp. 7--10.

\bibitem{guo2017deepfm}
H.~Guo, R.~Tang, Y.~Ye, Z.~Li, and X.~He, ``{D}eep{FM}: {A}
  factorization-machine based neural network for {CTR} prediction,'' in
  \emph{{IJCAI}}, 2017, pp. 1725--1731.

\bibitem{he2017nfm}
X.~He and T.-S. Chua, ``Neural factorization machines for sparse predictive
  analytics,'' in \emph{SIGIR}, 2017, pp. 355--364.

\bibitem{zhao2017gb}
Q.~Zhao, Y.~Shi, and L.~Hong, ``{GB-CENT}: Gradient boosted categorical
  embedding and numerical trees,'' in \emph{WWW}, 2017, pp. 1311--1319.

\bibitem{zhu2017deep}
J.~Zhu, Y.~Shan, J.~Mao, D.~Yu, H.~Rahmanian, and Y.~Zhang, ``Deep embedding
  forest: Forest-based serving with deep embedding features,'' in \emph{KDD},
  2017, pp. 1703--1711.

\bibitem{wang2018tem}
X.~Wang, X.~He, F.~Feng, L.~Nie, and T.-S. Chua, ``{TEM}: Tree-enhanced
  embedding model for explainable recommendation,'' in \emph{WWW}, 2018, pp.
  1543--1552.

\bibitem{Shi2019NeuACF}
C.~Shi, X.~Han, S.~Li, X.~Wang, S.~Wang, J.~Du, and P.~Yu, ``Deep collaborative
  filtering with multi-aspect information in heterogeneous networks,''
  \emph{{IEEE} Trans. Knowl. Data Eng.}, 2019.

\bibitem{he2018outer}
X.~He, X.~Du, X.~Wang, F.~Tian, J.~Tang, and T.-S. Chua, ``Outer product-based
  neural collaborative filtering,'' in \emph{IJCAI}, 2018, pp. 2227--2233.

\bibitem{Bahdanau2015nmt}
D.~Bahdanau, K.~Cho, and Y.~Bengio, ``Neural machine translation by jointly
  learning to align and translate,'' in \emph{{ICLR}}, 2015.

\bibitem{He2018NAIS}
X.~He, Z.~He, J.~Song, Z.~Liu, Y.-G. Jiang, and T.-S. Chua, ``{NAIS:} {N}eural
  attentive item similarity model for recommendation,'' \emph{{IEEE} Trans.
  Knowl. Data Eng.}, vol.~30, no.~12, pp. 2354--2366, 2018.

\bibitem{tay2018self}
Y.~Tay, S.~Zhang, L.~A. Tuan, and S.~C. Hui, ``Self-attentive neural
  collaborative filtering,'' \emph{arXiv preprint arXiv:1806.06446}, 2018.

\bibitem{Cao2019seagr}
D.~{Cao}, X.~{He}, L.~{Miao}, G.~{Xiao}, H.~{Chen}, and J.~{Xu},
  ``Social-enhanced attentive group recommendation,'' \emph{{IEEE} Trans.
  Knowl. Data Eng.}, 2019.

\bibitem{Xi2019bpam}
W.-D. Xi, L.~Huang, C.-D. Wang, Y.-Y. Zheng, and J.~Lai, ``{BPAM:}
  {R}ecommendation based on {BP} neural network with attention mechanism,'' in
  \emph{{IJCAI}}, 2019, pp. 3905--3911.

\bibitem{Chen2017acf}
J.~Chen, H.~Zhang, X.~He, L.~Nie, W.~Liu, and T.-S. Chua, ``{A}ttentive
  {C}ollaborative {F}iltering: {M}ultimedia recommendation with item- and
  component-level attention,'' in \emph{{SIGIR}}, 2017, pp. 335--344.

\bibitem{Cheng2018a3ncf}
Z.~Cheng, Y.~Ding, X.~He, L.~Zhu, X.~Song, and M.~Kankanhalli, ``{A3NCF}: {A}n
  adaptive aspect attention model for rating prediction,'' in \emph{IJCAI},
  2018, pp. 3748--3754.

\bibitem{Xiao2017afm}
J.~Xiao, H.~Ye, X.~He, H.~Zhang, F.~Wu, and T.-S. Chua, ``Attentional
  factorization machines: Learning the weight of feature interactions via
  attention networks,'' in \emph{IJCAI}, 2017, pp. 3119--3125.

\bibitem{Zhou2017atrank}
C.~Zhou, J.~Bai, J.~Song, X.~Liu, Z.~Zhao, X.~Chen, and J.~Gao, ``{ATR}ank:
  {A}n attention-based user behavior modeling framework for recommendation,''
  in \emph{{AAAI}}, 2018, pp. 4564--4571.

\bibitem{wu2016collaborative}
Y.~Wu, C.~DuBois, A.~X. Zheng, and M.~Ester, ``Collaborative denoising
  auto-encoders for top-n recommender systems,'' in \emph{WSDM}, 2016, pp.
  153--162.

\bibitem{pan2008one}
R.~Pan, Y.~Zhou, B.~Cao, N.~N. Liu, R.~Lukose, M.~Scholz, and Q.~Yang,
  ``One-class collaborative filtering,'' in \emph{ICDM}, 2008, pp. 502--511.

\bibitem{raffel2015feed}
C.~Raffel and D.~P.~W. Ellis, ``Feed-forward networks with attention can solve
  some long-term memory problems,'' \emph{CoRR}, vol. abs/1512.08756, 2015.

\bibitem{Zhou2018DIN}
G.~Zhou, X.~Zhu, C.~Song, Y.~Fan, H.~Zhu, X.~Ma, Y.~Yan, J.~Jin, H.~Li, and
  K.~Gai, ``Deep interest network for click-through rate prediction,'' in
  \emph{KDD}, 2018, pp. 1059--1068.

\bibitem{goh1995back}
A.~T. Goh, ``Back-propagation neural networks for modeling complex systems,''
  \emph{Artificial Intelligence in Engineering}, vol.~9, no.~3, pp. 143--151,
  1995.

\bibitem{he2016fast}
X.~He, H.~Zhang, M.-Y. Kan, and T.-S. Chua, ``Fast matrix factorization for
  online recommendation with implicit feedback,'' in \emph{SIGIR}, 2016, pp.
  549--558.

\bibitem{james2013cross}
J.~D. McCaffrey, ``Why you should use cross-entropy error instead of
  classification error or mean squared error for neural network classifier
  training,'' 2013, online; accessed 4-July-2018.

\bibitem{kingma2014adam}
D.~P. Kingma and J.~Ba, ``Adam: {A} method for stochastic optimization,'' in
  \emph{{ICLR}}, 2015.

\bibitem{erhan2010does}
D.~Erhan, Y.~Bengio, A.~C. Courville, P.-A. Manzagol, P.~Vincent, and
  S.~Bengio, ``Why does unsupervised pre-training help deep learning?''
  \emph{J. Mach. Learn. Res.}, vol.~11, pp. 625--660, 2010.

\bibitem{sarwar2001item}
B.~Sarwar, G.~Karypis, J.~Konstan, and J.~Riedl, ``Item-based collaborative
  filtering recommendation algorithms,'' in \emph{WWW}, 2001, pp. 285--295.

\end{thebibliography}


% Generated by IEEEtran.bst, version: 1.14 (2015/08/26)
\bibliographystyle{IEEEtran}

\end{document}